\PassOptionsToPackage{dvipsnames,table}{xcolor}
\documentclass[10pt]{article} 
\usepackage[preprint]{tmlr}
\usepackage[colorlinks=true,citecolor=perfblue,linkcolor=somecolor]{hyperref} 
\definecolor{somecolor}{RGB}{178,24,43}
\definecolor{perfblue}{RGB}{64, 114, 175}
\usepackage[page,toc,titletoc,title]{appendix}
\usepackage{graphicx} 
\usepackage{wrapfig}
\usepackage{amsmath,amssymb,amsthm}
\usepackage{subcaption}
\usepackage[utf8]{inputenc} 
\usepackage[T1]{fontenc}    
\usepackage{url}            
\usepackage{makecell}
\usepackage{booktabs}       
\usepackage{amsfonts}       
\usepackage{nicefrac}       
\usepackage{microtype}      
\usepackage{xspace}
\usepackage{url}
\usepackage{fontawesome5}
\usepackage{mdframed}
\usepackage{tikz}
\usepackage{multirow}
\RequirePackage[export]{adjustbox}
\usepackage{pifont}
\usepackage{cleveref}
\usepackage{algorithm}
\usepackage{algpseudocode}
\usepackage{tablefootnote}

\title{\themodel: Long-context Protein Language Modeling Using Bidirectional Mamba with Shared Projection Layers}

\author{\name Yingheng Wang\thanks{This work was completed while the author was an intern at Amazon.} \email yw2349@cornell.edu \\
      \addr Department of Computer Science \\
      Cornell University
      \AND
      \name Zichen Wang\thanks{Corresponding author.} \email zichewan@amazon.com \\
      \addr Amazon
      \AND
      \name Gil Sadeh \email gilsadeh@amazon.com \\
      \addr Amazon
      \AND
      \name Luca Zancato \email zancato@amazon.com \\
      \addr Amazon
      \AND
      \name Alessandro Achille \email aachille@amazon.com \\
      \addr Amazon
      \AND
      \name George Karypis \email gkarypis@amazon.com \\
      \addr Amazon
      \AND
      \name Huzefa Rangwala\thanks{Huzefa Rangwala is on LOA as a Professor of Computer Science at George Mason University. This paper describes work performed at Amazon.} \email rhuzefa@amazon.com \\
      \addr Amazon
}

\definecolor{mark}{RGB}{214, 235, 245}

\newcommand{\themodel}{\textbf{\texttt{LC-PLM}}\xspace}
\newcommand{\themodelg}{\textbf{\texttt{LC-PLM-G}}\xspace}
\newcommand{\themodelb}{\textbf{\texttt{BiMamba}}\xspace}
\newcommand{\themodels}{\textbf{\texttt{BiMamba-S}}\xspace}

\begin{document}

\maketitle
\begin{abstract}
Self-supervised training of language models (LMs) has seen great success for protein sequences in learning meaningful representations and for generative drug design. 
Most protein LMs are based on the Transformer architecture trained on individual proteins with short context lengths. Such protein LMs cannot extrapolate to longer proteins and protein complexes well. They also fail to account for the underlying biological mechanisms carried out by biomolecular interactions and dynamics i.e., proteins often interact with other proteins, molecules, and pathways in complex biological systems. 
In this work, we propose \themodel based on an alternative protein LM architecture, \themodels, built upon selective structured state-space models, to learn high-quality universal protein representations at the amino acid token level using masked language modeling. We also introduce its graph-contextual variant, \themodelg, which contextualizes protein-protein interaction (PPI) graphs for a second stage of training. \themodel demonstrates favorable neural scaling laws, better length extrapolation capability, and up to 30\% and 16\% improvements on protein downstream tasks compared to Transformer-based ESM-2 when trained with 100B and 1T tokens, respectively. \themodelg further trained within the context of PPI graphs shows promising results on protein structure and function prediction tasks. Our study demonstrates the benefit of increasing the context size with computationally efficient LM architecture (e.g., structured state space models) in learning universal protein representations and incorporating molecular interaction contexts contained in biological graphs.

\end{abstract}

\section{Introduction}

Most biological sequences are derived from genomes, which are long DNA sequences: human chromosomes range from 50 to 300 million base pairs. The protein-coding regions, which can be considered as the translated substrings of the genome, are relatively shorter (the majority are < 3,000 amino acids), albeit with a few exceptions, such as Titin, composed of 34K amino acids. The prevalent protein language models (pLMs), e.g. ESM-2 \citep{lin2023evolutionary}, choose 1024 as the context length as it fits 97.4\% of proteins. However, it does not natively support tasks that require long-range context windows to reason over multiple related sequences, such as genomic interactions, protein-protein interactions (PPI), protein function prediction, and 3D structure prediction of long proteins and protein complexes. Another challenge for modeling long-range biological contexts lies in their non-sequential nature. For instance, the useful context for genomic interactions and PPIs often span across regions from different chromosomes, and capturing information within an LM of such interactions usually requires biomedical knowledge graphs for good performance on these tasks \citep{kovacs2019network, sousa2024explaining}.

Large LMs including those trained on protein sequences, are predominantly based on the Transformer \citep{vaswani2017attention} with multi-head attention. Despite its state-of-the-art performance on virtually all types of data modalities (texts, vision, audio, etc.), it suffers from quadratic time and space complexity due to the lengths of the input sequences. 
Additionally, transformer models are known to have poor length extrapolation quality and do not achieve the same level of performance when evaluated on sequences longer than seen during pretraining. Recent work in alternative architectures such as convolutional (e.g.~Hyena \citep{poli2023hyena}) and selective structured state space models (SSMs) (e.g.~Mamba \citep{gu2024mambalineartimesequencemodeling}) have demonstrated competitive performance and preferable scaling properties on long context compared to Transformers and extensions including linear attention approximation variants \citep{katharopoulos2020transformers,zhai2021attention,peng2023rwkv}.
Although recent studies have leveraged these novel architectures to train LMs for DNA sequences \citep{nguyen2024hyenadna,nguyen2024sequence,schiff2024caduceus}, studies examining their feasibility as protein LMs are limited. There is also a research gap on how to effectively leverage the long-context capability of these architectures to model graphs of sequences i.e., how to leverage PPI graphs to improve LM's ability to reason across interacting (related) proteins. 

\begin{wrapfigure}{r}{0.57\textwidth} 
\vspace{-12pt}
  \centering
  \includegraphics[width=\linewidth]{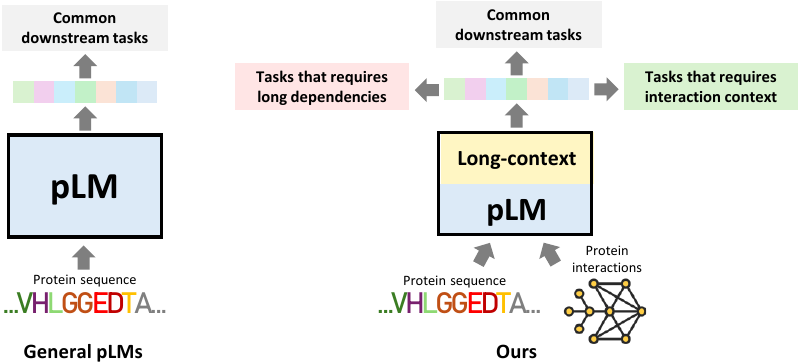}
  \vspace{-17pt}
  \caption{Our model enables long-context capability, length extrapolation ability, better neural scaling law, and interaction context-aware inference.}
  \label{fig:abs}
  \vspace{-15pt}
\end{wrapfigure}

In this work, we explore alternative architectures based on Mamba to improve the long-context capability of pLMs. We train a long-context pLM (\themodel) using bidirectional Mamba with shared projection layers (\themodels) on protein sequences from \texttt{UniRef50} with masked language modeling (MLM) objective. Results show favorable neural scaling laws, length extrapolation properties on \texttt{UniRef90}, and better downstream task performance on \texttt{TAPE} \citep{rao2019evaluating} and \texttt{ProteinGym} \citep{notin2024proteingym} than its Transformer counterpart, namely ESM-2. Such long-context and length extrapolation properties facilitate and improve structure prediction of long proteins and protein complexes from \texttt{CASP14}, \texttt{CASP15-multimers}, and \texttt{Benchmark2}. Next, we train a graph-contextualized variant \themodelg, which uses a proposed novel second-stage training strategy to leverage the long-context capabilities of \themodels to encode useful information from interaction graphs (e.g., PPI). Trained on sampled random walks that are composed of sequences of proteins, \themodelg improves performance on remote homology prediction, node-level protein function prediction (\texttt{ogbn-proteins}), and link-level PPI prediction (\texttt{ogbl-ppa}) \citep{hu2020open}. 
Our contributions can be summarized into three folds as follows:
\begin{itemize}
    \item We develop a long-context pLM (\themodel) with an alternative architecture based on a more sample \& compute-efficient bidirectional Mamba architecture with shared projection layers (\themodels) pretrained on \texttt{UniRef50} with MLM objective.
    \item We demonstrate that \themodel has improved length extrapolation capabilities and favorable scaling laws, and outperforms ESM-2 by up to 30\% and 16\% on various downstream tasks (e.g.~protein structure prediction (\texttt{CASP15-multimers}, \texttt{CASP14}, \texttt{Benchmark2}), tasks in \texttt{TAPE} and \texttt{ProteinGym}) when trained with 100B and 1T tokens, respectively.
    \item To encode biological interaction information, we propose a novel second-stage training based on random walks on graphs to extend the long-context capabilities of \themodel that leverages the PPI graph context. We demonstrate its effectiveness in capturing graph-contextual information on predicting remote homology (\texttt{TAPE}), protein functions (\texttt{ogbn-proteins}), and PPI links (\texttt{ogbl-ppa}). 
\end{itemize}

\section{Related Works}

\subsection{Long-context LMs and State Space Models}

Since their introduction, Transformers \citep{vaswani2017attention} with multi-head attention have been successfully applied in many different applications in natural language and computer vision. 
However, while being relatively straightforward to scale the number of parameters, Transformer models have a quadratic dependence on the context length during training and are linear at inference time, making them expensive to scale to long context. 
Alternative to Transformers, Recurrent Neural Networks (RNNs) \citep{hochreiter1991untersuchungen, bengioRNN, LSTMs1997} scale more favorably with context length and have linear dependency at training time and constant at inference time. However, generic non-linear RNNs cannot be parallelized on modern hardware due to the sequential nature of their gradient update rule \citep{bengioRNN}. 

To improve RNNs scalability on modern hardware, recent works on SSMs \citep{gu2021efficiently, fu2022hungry, gu2024mambalineartimesequencemodeling} propose to linearize RNNs dynamics and use efficient hardware-aware algorithms. A notable example is Mamba \citep{gu2024mambalineartimesequencemodeling}, which leverages the associative scan to efficiently process arbitrarily long sequences in linear time, and Mamba-2 \citep{dao2024transformers} that greatly improves over Mamba by implementing SSM layers using structured matrix multiplications to better leverage modern Tensor cores. 

To further harvest the benefits of SSM and Transformer primitives, hybrid models have been proposed in \cite{zancato2024b, lieber2024jamba, arora2024simple, de2024griffinmixinggatedlinear, botev2024recurrentgemma, waleffe2024empirical}.
There are also efforts trying to extend Mamba models to graph data \citep{wang2024graphmambalongrangegraphsequence, behrouz2024graphmambalearninggraphs}. However, unlike our \themodelg, which learns token-level protein representations within graph context from the graph of sequences, they focus on learning node/graph-level representations that only work for generic graph tasks where nodes do not contain sequences (see \Cref{tab:sum})\footnote{We provide the concrete definitions in \Cref{app:cri} for all criterions used in \Cref{tab:sum}.}.

\begin{table}[h!]
\centering
\begin{adjustbox}{width=0.99\linewidth, center}
\begin{tabular}{l c c c c c}
\toprule
\textbf{Method}  & \textbf{Universality} & \makecell{\textbf{Fine} \\ \textbf{granularity}} & \makecell{\textbf{Long-context capability} \\ \textit{Handleability} \& \textit{Performance} } & \makecell{\textbf{Graph} \\ \textbf{context}} & \makecell{\textbf{Large-scale} \\ \textbf{model}} \\
\midrule
ProtGPT (\citeyear{ferruz2022protgpt2}) & \ding{51} & \ding{51} & \ding{55} ~~~~~~~~~~~~~~~~~~~~~ \ding{55} & \ding{55} & \ding{55} \\
ESM-2 (\citeyear{lin2023evolutionary}) & \ding{51} & \ding{51} & \ding{55} ~~~~~~~~~~~~~~~~~~~~~ \ding{55} & \ding{55} & \ding{51} \\
CARP (\citeyear{yang2024convolutions}) & \ding{51} & \ding{51} & \ding{51} ~~~~~~~~~~~~~~~~~~~~~ \ding{55} & \ding{55} & \ding{51} \\
ProtHyena (\citeyear{zhang2024prothyena}) & \ding{51} & \ding{51} & \ding{51} ~~~~~~~~~~~~~~~~~~~~~ \ding{55} & \ding{55} & \ding{55} \\
PoET (\citeyear{truong2024poet}) & \ding{55} & \ding{51} & \ding{55} ~~~~~~~~~~~~~~~~~~~~~ \ding{55} & \ding{55} & \ding{55} \\
ProtMamba (\citeyear{sgarbossa2024protmamba}) & \ding{55} & \ding{51} & \ding{55}\tablefootnote{ProtMamba used an unextrapolatable learnable positional encoding that sacrificed the long-context capability of Mamba. We discuss more in \Cref{app:cri}.}~~~~~~~~~~~~~~~~~~~~~ \ding{55} & \ding{55} & \ding{55} \\
PTM-Mamba (\citeyear{peng2024ptm}) & \ding{55} & \ding{51} & \ding{51} ~~~~~~~~~~~~~~~~~~~~~ \textbf{--} & \ding{55} & \ding{51} \\
\midrule
Graph-Mamba (\citeyear{wang2024graphmambalongrangegraphsequence}) & \ding{55} & \ding{55} & \ding{51} ~~~~~~~~~~~~~~~~~~~~~ \textbf{--} & \ding{51} & \ding{55} \\
GMN (\citeyear{behrouz2024graphmambalearninggraphs}) & \ding{55} & \ding{55} & \ding{51} ~~~~~~~~~~~~~~~~~~~~~ \textbf{--} & \ding{51} & \ding{55} \\
\midrule
\themodel & \ding{51} & \ding{51} & \ding{51} ~~~~~~~~~~~~~~~~~~~~~ \ding{51} & \ding{55} & \ding{51} \\
\themodelg & \ding{51} & \ding{51} & \ding{51} ~~~~~~~~~~~~~~~~~~~~~ \ding{51} & \ding{51} & \ding{51} \\
\bottomrule
\end{tabular}
\end{adjustbox}
\caption{Comparison of \themodel and \themodelg to other protein LMs and graph SSMs in terms of enabling universal representations, AA token-level fine granularity, long-context capability, graph contextual information, large model size, and a large number of pretrained tokens.}
\label{tab:sum}
\end{table}

\subsection{Long-context LMs for biological sequences}

To model the long-range interactions without sacrificing single nucleotide level resolution, long-context capable LM architectures have been developed for DNA sequences, including HyenaDNA \citep{nguyen2024hyenadna}, Evo \citep{nguyen2024sequence}, and Caduceus \citep{schiff2024caduceus}. These studies have shown that alternative architectures based on SSMs exhibit better scaling laws than Transformers on genomic data and DNA-specific tasks.  
Protein sequence LMs with alternative architectures have also been explored to improve computational efficiency and enable the modeling of longer protein sequences. For instance, CARP \citep{yang2024convolutions} is a protein LM with dilated convolution layers. Pretrained with MLM objective, CARP achieved comparable pretraining scaling properties with its Transformer counterpart ESM-1b \citep{rives2021biological} and scales better on long sequences. ProtHyena \citep{zhang2024prothyena} is a small 1.6M parameter decoder-only LM based on the Hyena operator pretrained on protein sequences and has demonstrated some improvement over ProtGPT \citep{ferruz2022protgpt2} with comparable model sizes. 

Some works exploit the long-context capability of LMs to model sets of homologous protein sequences such as those in multiple sequence alignment (MSA), which further organize a set of protein sequences by aligning evolutionary conserved amino acids across the set of sequences. PoET \citep{truong2024poet} proposed a tiered variant of Transformer to model the invariant relationships between multiple protein sequences from MSAs, whereas ProtMamba \citep{sgarbossa2024protmamba} trains a Mamba-based protein LM using concatenated sequences from MSAs with causal language modeling and infilling objective. PTM-Mamba \citep{peng2024ptm} addresses post-translational modifications (PTM) of protein sequences introducing PTM tokens to amino acid tokens and subsequently trains a bidirectional Mamba model with these PTM tokens. We provide additional discussion on other pLMs and related variants in \Cref{app:plm}. 

Instead of training protein sequences from very specific types of data like MSAs or PTMs, we emphasize that \textbf{our work} focuses on long-context modeling of individual protein sequences and related protein sequences within biomedical graphs, which learns universal AA token-level protein representations that are more generalizable and can encode information from biological interactions (see \Cref{tab:sum} for a detailed comparison).

\subsection{Protein LMs trained on graphs}

Graphs are ubiquitous in biomedical domains as they are suitable for organizing and representing complex biological systems, such as gene regulatory networks and PPI graphs \citep{wang2022graph}. The relationships among proteins embedded in biomedical graphs have also been used to train pLMs. The common strategies for incorporating graph information include pretraining LMs with graph-specific objectives in addition to self-supervised LM objectives. The graph-specific objective can be link-prediction on homogeneous graphs \citep{yasunaga2022linkbert,mcdermott2023structure}, knowledge graph embedding (KGE) objectives \citep{zhang2022ontoprotein} or contrastive loss \citep{wang2023biobridge} on heterogeneous graphs. One limitation of such approaches is the inability to jointly model the implicit token-wise interactions beyond a pair of sequences. After all, link-prediction and KGE only take two sequences as input. In our work, we use homogeneous PPI graphs and exploit the long-context capability of SSM-based LM to model token-wise interactions beyond two sequences. 

\section{Preliminaries}

\paragraph{Structured State Space Models}
Modern Structured SSMs are derived from first-order differential equations that map the input sequence $x(t)$ to the output sequence $y(t)$ through hidden state $h(t)$:
\begin{align}
    \mathbf{h'}(t) &= \mathbf{A} \mathbf{h}(t) + \mathbf{B} x(t), \quad y(t) = \mathbf{C} \mathbf{h}(t) \label{eq:ssm1}
\end{align}
where $\mathbf{A} \in \mathbb{R}^{N \times N}$, $\mathbf{B} \in \mathbb{R}^{N \times D}$ and $\mathbf{C} \in \mathbb{R}^{D \times N}$. The variables $N$ and $D$ refer to the state dimension and the (expanded) input dimension respectively. The continuous dynamical system characterized by $\mathbf{A}$, $\mathbf{B}$ can be discretized to $\bar{\mathbf{A}}, \bar{\mathbf{B}}$ by zero-order holding and time sampling at intervals of $\Delta$, defined as follows:
\begin{align}
    \bar{\mathbf{A}} &= \exp(\Delta \mathbf{A}), \quad \bar{\mathbf{B}} = (\Delta \mathbf{A})^{-1} (\exp(\Delta \mathbf{A}) - \mathbf{I}) \cdot \Delta \mathbf{B}. \label{eq:ssm2}
\end{align}
The formula of a discretized SSM can then be written as:
\begin{align}
    \mathbf{h}_k &= \bar{\mathbf{A}} \mathbf{h}_{k-1} + \bar{\mathbf{B}} x_k, \quad y_k = \mathbf{C} \mathbf{h}_k \label{eq:ssm3}
\end{align}

The main benefit of discretized SSMs \citep{gu2021efficiently} over their continuous counterpart is that they can be trained efficiently using their parallel convolutional representation and can be efficiently deployed at inference time with their recurrent form. However, the ability to model long-range interactions of SSMs is limited by the impulse response of the discrete dynamical system they implement, the S4 model \citep{gu2020hippo, gu2021efficiently} mitigates such limitation by introducing the HIPPO Matrix to the initialization of $\mathbf{A}$.

\paragraph{Selection Mechanism and Mamba}

The main limitation of the SSMs described so far is that they cannot model complex input-varying interactions across the sequence dimension. Thus, Mamba \citep{gu2024mambalineartimesequencemodeling} parameterizes the matrices $\mathbf{B}$, $\mathbf{C}$ and $\Delta$ in an input-dependent (data-driven) manner, introducing a selection mechanism into the S4 model. However, introducing such data dependency makes the parallelizable convolutional representation unfeasible, hence, Mamba uses a novel hardware-aware parallel computing algorithm (based on the associative scan) to ensure the efficient training of the model and leading to linear computational complexity and outstanding capabilities in modeling long-term dependencies.
A Mamba model (\textit{selective} SSM) that enables dependence of the parameters $\mathbf{B}$, $\mathbf{C}$ and $\Delta$ on the input $x_t$ can be formulated as:
\begin{align}
    \mathbf{B}_t &= \text{Linear}_\mathbf{B}(x_t) \quad \mathbf{C}_t = \text{Linear}_\mathbf{C}(x_t) \\
    \Delta_t &= \text{softplus}(\text{Linear}_\Delta(x_t)), \label{eq:selective_ssm}
\end{align}
where $\text{Linear}(\cdot)$ represents a linear projection and $\text{softplus}(\cdot) = \log(1 + \exp(\cdot))$.

\section{\themodel: Long Context Protein Language Model}

In this section, we first introduce the design choice of using bidirectional Mamba (\themodelb) with shared projection layers (\themodels) for building up the model architecture of \themodel, and then we discuss how we develop the two-stage training recipe to obtain high-quality universal protein representations using MLM and encode biologically meaningful interaction information with a novel graph context-aware training approach.

\subsection{\themodels: Bidirectional Mamba with Shared Projection Layers}

\themodelb is an extension from standard Mamba block and has been applied in various domains, e.g. time-series forecasting \citep{liang2024bi}, audio representation learning \citep{erol2024audio}, visual representation learning \citep{zhu2024vision}, DNA modeling \citep{schiff2024caduceus}, and graph learning \citep{behrouz2024graphmambalearninggraphs}. The following reasons suggest we consider \themodelb as the design choice: 
(i) Mamba is good at {\emph{capturing long-range dependencies}} and {\emph{extrapolating on longer sequences}}, which benefit a lot of downstream tasks on protein complexes and PPI graphs.
(ii) standard Mamba only does unidirectional (associative) scans for causal sequence modeling. To {\emph{perform MLM to learn high-quality universal protein representations}}, we introduce a modified bidirectional scan to capture information from both ends.

In general, the $l$-th \themodelb block takes in an input sequence of tokens $\mathbf{T}_{l-1} \in \mathbb{R}^{B \times S \times D}$ and output $\mathbf{T}_{l} \in \mathbb{R}^{B \times S \times D}$ where $B, S, D$ represent the batch size, the input dimension, and the hidden state dimension. Then a residual connection adds the input and output together to get $\mathbf{T}_l$. After going through $L \times$ \themodelb blocks, the output $\mathbf{T}_L$ will be normalized first and then fed into a prediction head to get final scores. This procedure can be formulated as follows:
\begin{align}
    &\mathbf{T}_l = \text{BiMamba}\left( \mathbf{T}_{l-1} \right) + \mathbf{T}_{l-1} \\ &\hat{p} = \text{PredictionHead}\left( \text{Norm} \left(\mathbf{T}_{L}\right) \right)
\end{align}

Specifically, in one \themodelb block, the input sequence $\mathbf{T}_{l-1}$ and the flipped $\hat{\mathbf{T}}_{l-1}$ will be first normalized and then linearly projected to $\mathbf{X}_{l-1} \in \mathbb{R}^{B \times S \times E}$ and $\mathbf{Z}_{l-1} \in \mathbb{R}^{B \times S \times E}$. $\mathbf{X}_{l-1}$ and the flipped $\hat{\mathbf{X}}_{l-1}$ will be fed into the forward and inverse Mamba block respectively for a bidirectional scan. In each Mamba block, $\mathbf{X}_{l-1}$ and $\hat{\mathbf{X}}_{l-1}$ will be first passed through a 1-D convolution layer and a SiLU activation \citep{nwankpa2018activation}, and then linearly projected to $ \mathbf{B}_{l-1} \in \mathbb{R}^{B \times S \times N}, \mathbf{C}_{l-1} \in \mathbb{R}^{B \times S \times N}, \Delta_{l-1} \in \mathbb{R}^{B \times S \times E}$ and $ \hat{\mathbf{B}}_{l-1}, \hat{\mathbf{C}}_{l-1}, \hat{\Delta}_{l-1}$, where $\Delta_{l-1}$ and $\hat{\Delta}_{l-1}$ transform $\mathbf{A}_{l-1}, \mathbf{B}_{l-1}$ and $\hat{\mathbf{A}}_{l-1}, \hat{\mathbf{B}}_{l-1}$ to $\bar{\mathbf{A}}_{l-1} \in \mathbb{R}^{B \times S \times E \times N}, \bar{\mathbf{B}}_{l-1} \in \mathbb{R}^{B \times S \times E \times N}$ and $\hat{\bar{\mathbf{A}}}_{l-1}, \hat{\bar{\mathbf{B}}}_{l-1}$. A standard SSM block will be then applied to obtain $\mathbf{Y}_{l-1} \in \mathbb{R}^{B \times S \times E}$ and $\hat{\mathbf{Y}}_{l-1}$, which later will be gated by $\mathbf{Z}_{l-1}$ and added together to get the candidate output. Lastly, a residual connection will be applied on a linear projection of the candidate output and input sequence $\mathbf{T}_{l-1}$ to get the final output $\mathbf{T}_l$. We provide an algorithmic block and a detailed breakdown to describe this computation process in \Cref{app:pseudo}.

\begin{wrapfigure}{r}{0.4\textwidth} 
    \vspace{-23pt}
    \centering
    \includegraphics[width=0.9\linewidth]{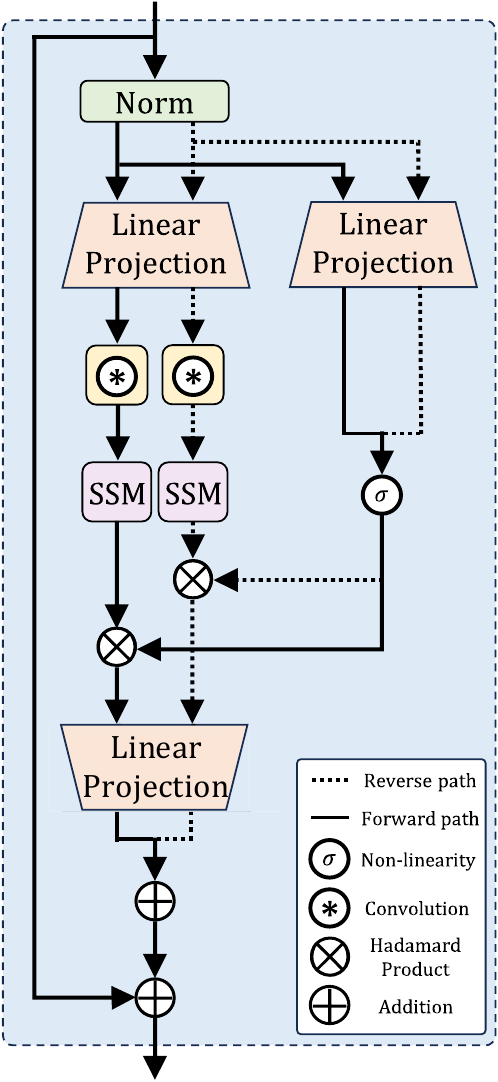}
    \vspace{-8pt}
    \caption{\themodels block. The forward and reverse modules share the linear projection layers. The normalized input will be reversed along the sequence dimension before being fed in. The output of the reversed will be flipped back and then added to the forward's output.}
    \label{fig:bimamba}
    \vspace{-70pt}
\end{wrapfigure}

\paragraph{Shared Projection Layers}
To explore a more efficient implementation of \themodelb, we propose to use the shared linear projection layers for the forward input $\mathbf{T}_{l-1}$ and the flipped $\hat{\mathbf{T}}_{l-1}$. This design choice helps make the entire model $2\times$ deeper with almost the same parameter counts \citep{schiff2024caduceus}. We refer to this building block as \themodels (illustrated in \Cref{fig:bimamba}). Note that this is different from the inner BiMamba block used in \cite{zhang2024mamba,zhu2024vision}, where they just flipped the linearly projected hidden states. We also find that, empirically, the deeper model using \themodels shows superiority in terms of sample \& compute efficiency (4.5\% improvement on evaluation loss) and performance gain on downstream tasks (an average of 4.1\% improvement on TM score of structure prediction) as we expected. More results and details are shown in \Cref{exp:weight}.

\paragraph{Untied Input \& Output Embeddings}
Notably, we opt to use untied input and output embeddings for the \themodels encoder. Empirically we find that untied embeddings yield better evaluation loss during MLM training compared to tied embeddings, despite the latter being the standard practice. This finding aligns with previous research \citep{gao2019representation,ethayarajh2019contextual}, which highlights that tying input and output embeddings leads to \textit{anisotropic} word embeddings in contextualized pretrained models, significantly constraining their expressiveness.

\subsection{Two-stage Training Recipe}
Our training procedure can be decomposed into two stages: (i) long-context protein language modeling and (ii) protein language modeling within graph contexts. The first stage will enforce \themodel to learn the universal token-level representations of individual proteins and the second stage will put the protein sequences into the related graph contexts and \themodelg will learn to capture biologically meaningful interaction information.

\paragraph{Long-context Protein Language Modeling}
\themodel is trained with \themodels on individual protein sequences where it can leverage the power of SSM modules to effectively capture long-range dependencies within sequences. Specifically, treating the protein sequences as a collection of amino acid (AA) tokens, the model learns fine granular and universal token-level representations using MLM, which can be generalized across different types of protein sequences. For the masking strategy, we follow BERT \citep{devlin2018bert} in which 15\% of AA tokens in a protein sequence will be "masked". Of the ‘masked’ tokens, 80\% are replaced with  \texttt{[MASK]}, 10\% are replaced with a random token from the vocabulary, and 10\% are left unchanged.

\begin{figure}[h]
\begin{center}
\includegraphics[width=1\textwidth]{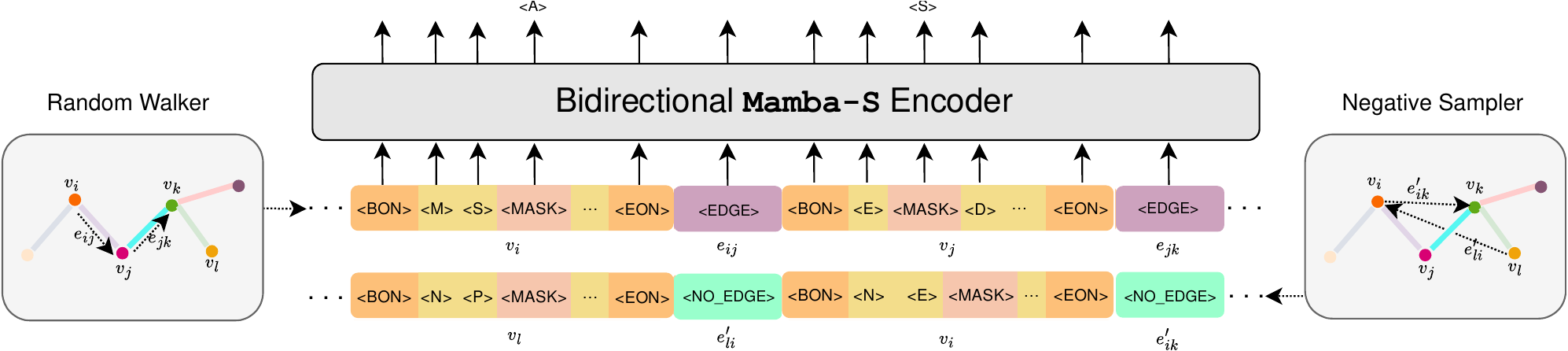}
\end{center}
\caption{The illustration of graph-contextual protein language modeling (\themodelg). The positive paths are sampled with random walks on the graph and the same number of negative paths are sampled from disconnected node pairs randomly. The sampled paths will be transformed into multi-protein sequences composed of AA tokens and special graph identifier tokens. The \themodels encoder is trained using MLM, the same as in the first-stage training.}
\label{fig:graphmamba}
\end{figure}

\paragraph{Graph-contextual Protein Language Modeling}

To encode biologically meaningful interaction information into protein representations, we propose the second-stage training within a graph context where a node represents an individual protein sequence and an edge indicates the existence of PPI. We refer to this graph-contextually trained model variant as \themodelg. \citet{wang2022graph} and \citet{behrouz2024graphmambalearninggraphs} propose to tokenize the graph into either flattened node sequences with prioritization strategy (e.g., node degree) or induced subgraphs. However, the former discards the graph topology information and the latter provides only 2-D tokens that cannot be used as the input of language models. Therefore, we propose to construct the graph-contextual input via random walks \citep{perozzi2014deepwalk, grover2016node2vec}, which can both effectively capture the graph topology and provide 1-D sequences. Consider an undirected, unweighted, PPI graph $\mathcal{G}=(V, E)$. Formally, a random walk of length $l$ can be simulated by
\begin{equation}
    P(n_i = v \mid n_{i-1} = u) = 
    \begin{cases} 
    \frac{\pi_{uv}}{Z} & \text{if } (v, u) \in E \\ 
    0 & \text{otherwise} 
    \end{cases}
\end{equation}
where $n_i$ denotes the $i$th node in the walk, $\pi_{uv}$ is the unnormalized transition probability between nodes $(u, v)$, and $Z$ is the normalizing constant. We also set two parameters $p$ and $q$ as in \cite{grover2016node2vec} to interpolate the behavior of random walker in between breath-first and depth-first search (see \Cref{app:sampling}). Then, the nodes in each random walk will be expanded as a sequence of proteins composed of AA tokens. We also sample a sequence of disconnected nodes of the same length $l$ as the negative paths. 

Although this gives us a principled way to form input multi-protein sequences for language models within graph context, the input still needs special identifiers to let the language model precept the graph topological information and be aware of which protein each AA token belongs to. Thus, we design four new tokens (\texttt{[BON]}, \texttt{[EON]}, \texttt{[EDGE]}, \texttt{[NO\_EDGE]}) to help encode such graph context information, where the first two indicate the begin and end of a node and the last two represent if there exists an edge. We provide a visual illustration of this graph-contextual training regime in \Cref{fig:graphmamba}.

\begin{figure}[htbp]
\centering
\begin{minipage}{0.48\textwidth}
    \centering
    \includegraphics[width=\textwidth]{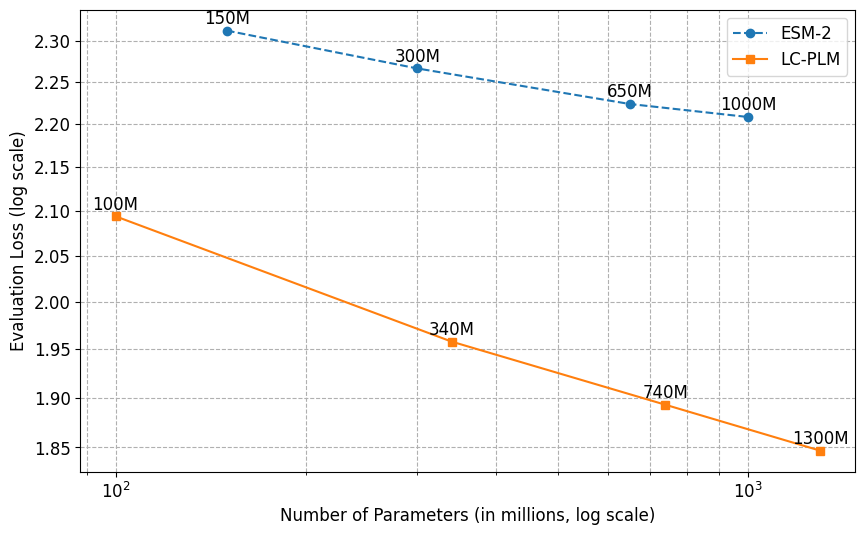}
    \caption{Evaluation loss across different model sizes for \themodel and ESM-2, showing that \themodel has a better scaling behavior when increasing the number of model parameters.}
    \label{fig:scaling}
\end{minipage}%
\hfill
\begin{minipage}{0.48\textwidth}
    \vspace{-15pt}
    \centering
    \includegraphics[width=\textwidth]{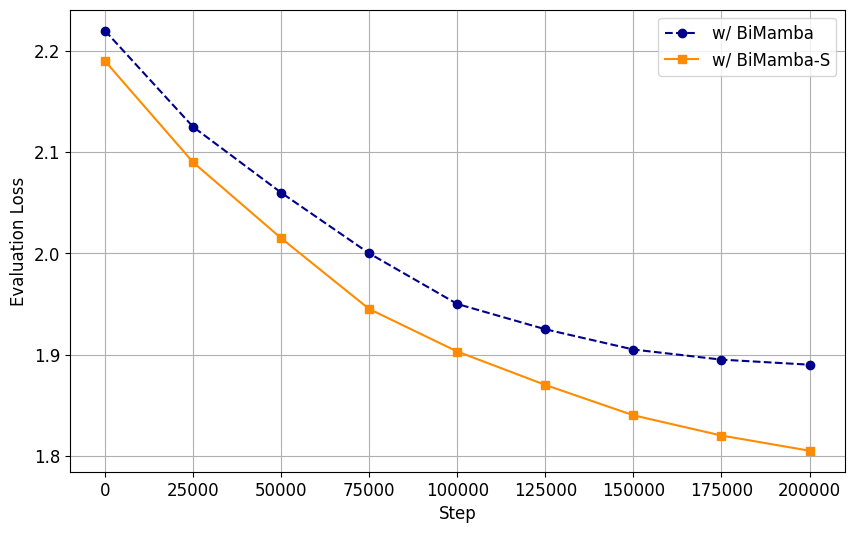}
    \caption{Evaluation loss comparison between \themodelb and \themodels over training steps.}
    \label{fig:weight}
\end{minipage}
\end{figure}

\section{Experiments}
 We conduct experiments to evaluate the effectiveness of \themodel and \themodelg and their building block \themodels. We will address the following research questions. \textbf{(RQ1)} What is the scaling behavior of \themodel? How does it compare with its Transformer-based counterpart ESM-2?
\textbf{(RQ2)} Does \themodel show stronger length extrapolation capability than ESM-2?
\textbf{(RQ3)} Will \themodels architecture be more effective in long-context protein language modeling?
\textbf{(RQ4)} Does long-range dependencies help with protein structure prediction?
\textbf{(RQ5)} Does \themodelg learn graph-contextual (relational) information?
\textbf{(RQ6)} Does \themodelg with biological interaction information learned in the second-stage training help with common downstream tasks?
\textbf{(RQ7)} Does \themodelg improve protein function prediction and link prediction on the PPI graph?
We provide the experimental setup, dataset description, and task definition in \Cref{app:data,app:exp}.

\subsection{\textbf{(RQ1)} Exploring the Scaling Law}
We train \themodel on 20B \texttt{UniRef90} sequences and evaluate it on a held-out set of 250K \texttt{UniRef90} sequences. We test four different model sizes for both \themodel and ESM-2. The model sizes for \themodel are 100M, 340M, 740M, and 1.3B parameters to accommodate \themodels architecture, while for ESM-2, they are 150M, 300M, 650M, and 1B. The results demonstrated that \themodel not only achieved better evaluation loss (average cross-entropy across all tokens) with a similar model size (with an average of 13.5\% improvement) compared to ESM-2 but also exhibited superior scaling behavior (sharper slope) when increasing the model size, as shown in \Cref{fig:scaling}. This aligns with the discovery in \cite{gu2024mambalineartimesequencemodeling} that Mamba has better neural scaling law compared to Transformers in language modeling. This may also be due to the useful long-range dependencies in protein sequences captured by \themodel and the deeper architecture achieved with \themodels.

\begin{wrapfigure}{r}{0.59\textwidth}
    \vspace{-20pt}
    \centering
    \includegraphics[width=1.02\linewidth]{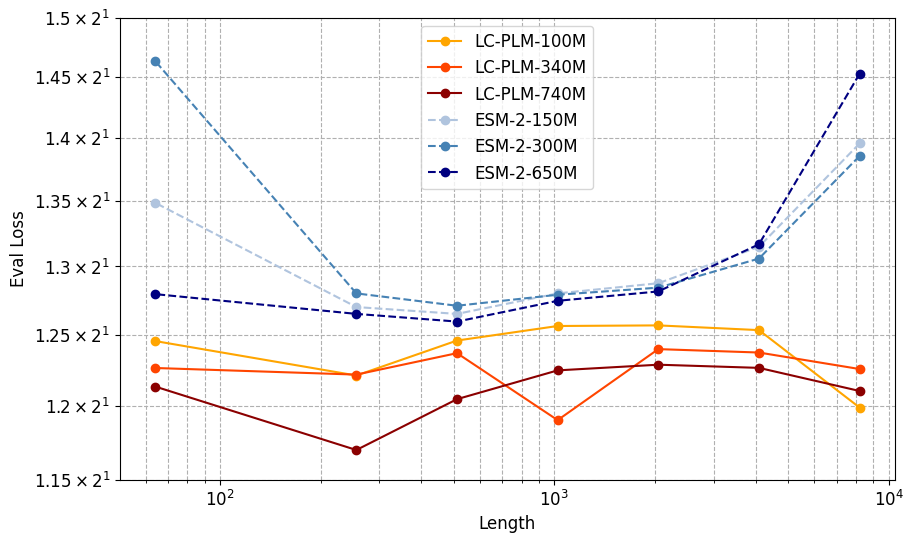}
    \vspace{-20pt}
    \caption{Length extrapolation results comparing \themodel versus ESM-2 on evaluation loss across different sequence lengths. \themodel can achieve consistent performance when extrapolating on longer sequences.}
    \label{fig:len_extra}
    \vspace{-15pt}
\end{wrapfigure}

\subsection{\textbf{(RQ2)} Length Extrapolation Evaluation}
We split the UniRef90 sequences into 7 bins w.r.t. the sequence length (i.e. 0-128, 128-256, 256-512, 512-1024, 1024-2048, 2048-4096, and 4096-8192). We train three sizes of \themodel (100M, 340M, 740M) and ESM-2 (150M, 300M, 650M) on the bin of 128-256 and then evaluate them on all bins (including a held-out set of 128-256). Our findings show that \themodel maintains low evaluation loss across sequence lengths, while ESM-2 struggles with both shorter and longer sequences, especially when the lengths are underrepresented in the training set. This concludes that \themodel can extrapolate better with length due to the stronger length extrapolation capability of \themodels \citep{gu2024mambalineartimesequencemodeling} compared to ESM-2, which uses RoPE \citep{su2024roformer} to extend context beyond pretraining. The results are shown in \Cref{fig:len_extra}.

\subsection{\textbf{(RQ3)} The Effectiveness of \themodels} \label{exp:weight}
Using shared linear projection layers in \themodels allows for $2\times$ deeper models with similar parameter counts. In our analysis, we compare the evaluation loss of our 740M model with \themodels and its \themodelb counterpart that halves the depth. The training set is \texttt{UniRef50} and the evaluation set is a held-out set of 250K \texttt{UniRef90} sequences, the same as in the scaling law experiments. Our results show that this parameter-efficient approach to increasing the model depth effectively improves evaluation loss by 4.5\%, as shown in \Cref{fig:weight}. We also verify the effectiveness of \themodels on structure prediction in \Cref{tab:sp}, where the deeper model improves by 6.7\% on \texttt{CASP15-multimers}, 4.6\% on \texttt{CASP14}, and 1.5\% on \texttt{Benchmark2}. This empirical evidence matches the theory that more hidden layers in deep neural networks can be benefited from more representation power gain, proposed in \citep{telgarsky2016benefits}. This also suggests a potentially better scaling strategy for training pLMs with the fixed parameter count, i.e. stacking more layers instead of having more hidden units.

\subsection{\textbf{(RQ4)} Protein Structure Prediction with LMFold}

We also evaluate \themodel’s ability to predict protein’s 3-D structures. It has been shown that protein LMs capture various levels of protein structure information \citep{rives2021biological, rao2020transformer,lin2023evolutionary}, despite being trained on primary sequences alone. Inspired by the ESMFold \citep{lin2023evolutionary} architecture, which uses the residue-level embeddings and, optionally, attention maps as features to predict the 3-D structures directly without MSA\footnote{We disable attention maps in our experiments since (i) there is no attention map in \themodels and (ii) ESMFold \citep{lin2023evolutionary} also demonstrate that attention maps provide no performance gain during training.}, we developed a protein folding model named LMFold, which generalizes ESMFold’s Folding Trunk and Structure Module to work with protein LMs with decoder-only and encoder-decoder architectures. Briefly, LMFold takes the residue-level embeddings for a given protein sequence as features to predict the all-atom coordinates of the protein structure. To further simplify LMFold, we only use 1 folding block of the structure module. Note that the goal of this task is not to develop the state-of-the-art protein folding model, but rather to quantify the potential of pretrained protein LMs for their learned structural information. To train LMFold, we use the \emph{Frame Aligned Point Error (FAPE)} and \emph{distogram} losses introduced in AlphaFold2 \citep{jumper2021highly}, as well as heads for predicting \emph{LDDT} and the \emph{pTM score}. We weigh these 4 loss terms using the default constants proposed in OpenFold \citep{ahdritz2024openfold}.

\begin{table}[ht!]
\centering
\begin{adjustbox}{width=0.99\linewidth, center}
\begin{tabular}{l c c c}
\toprule
\textbf{Model (\#Tokens trained)} & \textbf{CASP15-multimers} & \textbf{CASP14} & \textbf{Benchmark2} \\
\midrule
ESM-2-650M (100B)    & $0.3992 \pm 0.0418$  & $0.3403 \pm 0.0527$  & $0.4724 \pm 0.0407$  \\
\themodel-740M w/ \themodelb (100B)    & $0.4538 \pm 0.0151$  & $0.3827 \pm 0.0124$  & $0.6012 \pm 0.0101$  \\
\themodel-740M w/ \themodels (100B)  & \cellcolor{gray!15}${0.5012 \pm 0.01870}$  & ${0.4014 \pm 0.0190}$  & ${0.6128 \pm 0.0117}$  \\
\midrule
ProtMamba-public\tablefootnote{ProtMamba is hard to extrapolate on sequence > 2048 since they train with fixed-length positional encodings.}  & ${0.3561 \pm 0.0250}$  & ${0.3176 \pm 0.0219}$  & ${0.4431 \pm 0.0224}$  \\
ESM-2-650M-public (1T)\tablefootnote{The public ESM-2 model is provided for reference only. We highlight the best results for models trained with the same number of tokens and similar sizes. The tables below follow the same approach.}  & ${0.4750 \pm 0.0120}$  & \cellcolor{gray!15}${0.4300 \pm 0.0179}$  & \cellcolor{gray!15}${0.6491 \pm 0.0162}$  \\
\themodel-740M (1T)  & \cellcolor{gray!30}$\mathbf{0.5515 \pm 0.0254}$  & \cellcolor{gray!30}$\mathbf{0.4650 \pm 0.0215}$  & \cellcolor{gray!30}$\mathbf{0.7075 \pm 0.0248}$  \\
\bottomrule
\end{tabular}
\end{adjustbox}
\caption{Structure prediction performance (\emph{TM score}) on \texttt{CASP15-multimers}, \texttt{CASP14}, and \texttt{Benchmark2}. We perform 3 runs using different seeds and report the mean and standard deviation.}
\label{tab:sp}
\end{table}

For the training set, we down-sample 1.5\% of protein chains used in OpenFold \citep{ahdritz2024openfold}, leading to 7,872 chains, with at most 1 protein chain from each cluster. The aggressive down-sampling is supported by the fact that training a protein folding model with as few as 1,000 protein chains achieved a decent performance \citep{ahdritz2024openfold}. The down-sampled protein chains have lower than 40\% sequence identity to each other. We use 95\% and 5\% as data splitting for training and validation sets. For held-out test sets, we use \texttt{CASP15-multimers} (52 protein complexes), \texttt{CASP14} (37 protein structures), and \texttt{Benchmark2} (17 heterodimers structures) \citep{ghani2021improved}. We compare our 740M \themodel (with \themodelb or \themodels) against 650M ESM-2, all pretrained on 100B tokens from \texttt{UniRef50}. \themodel outperforms ESM-2 across all test sets by a large margin (20.8\% on \texttt{CASP15-multimers}, 17.6\% on \texttt{CASP14}, and 29.5\% on \texttt{Benchmark2}). \themodel also achieves comparable performance to 650M public ESM-2 model trained on 10$\times$ more tokens (1T), with 1.6\% improvement on \texttt{CASP15-multimers}. These results demonstrate the powerful long-context capability of \themodels on modeling longer proteins and protein complexes. This also suggests that, even for average-length protein sequences, long-range dependencies would be useful information and an important feature for protein structure prediction.

\begin{wrapfigure}{r}{0.62\textwidth}  
    \centering
    \begin{subfigure}[b]{0.49\linewidth}
        \centering
        \includegraphics[width=\linewidth]{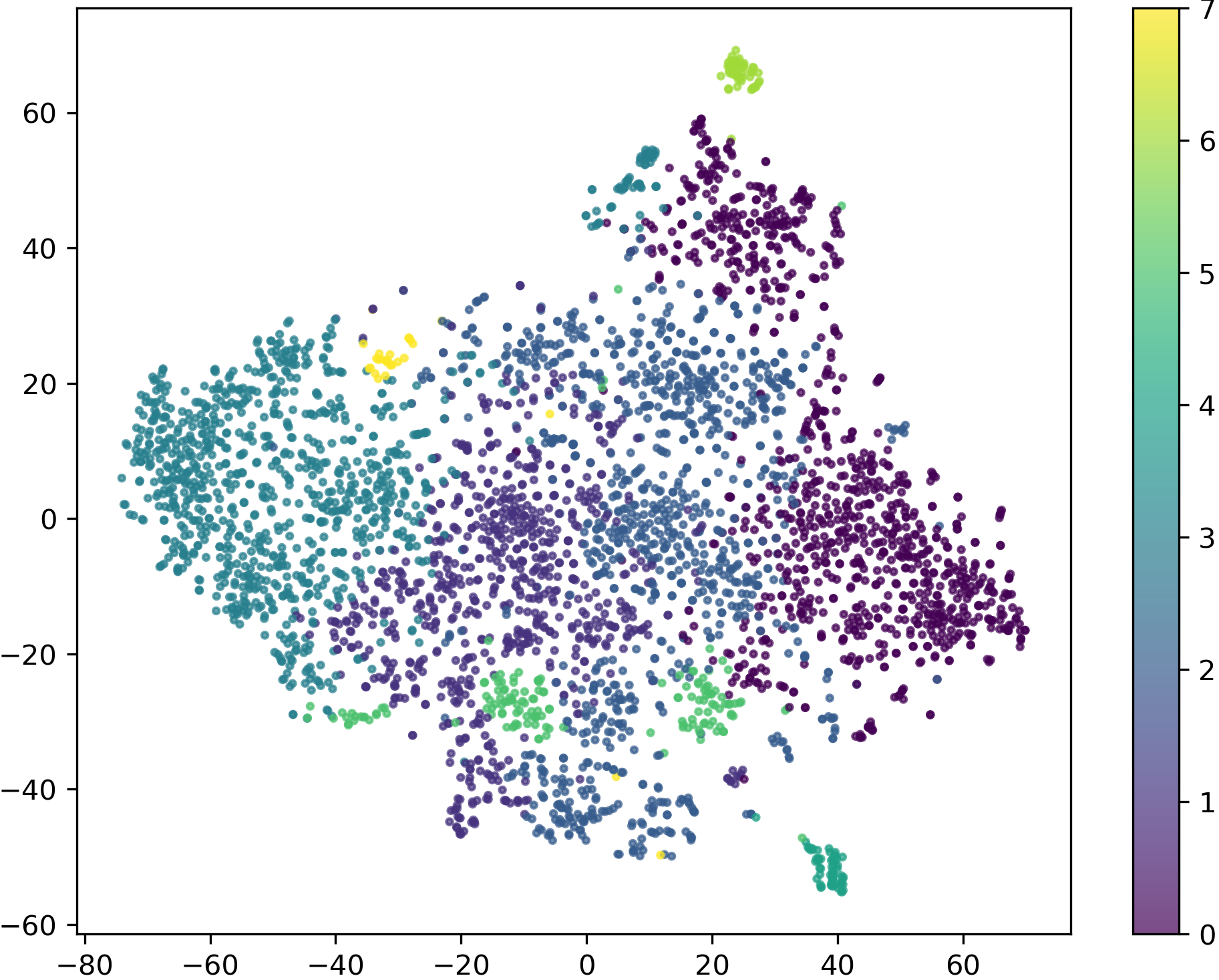}
        \caption{\themodel}
        \label{fig:tsne_avg_louvain}
    \end{subfigure}
    \hfill
    \begin{subfigure}[b]{0.49\linewidth}
        \centering
        \includegraphics[width=\linewidth]{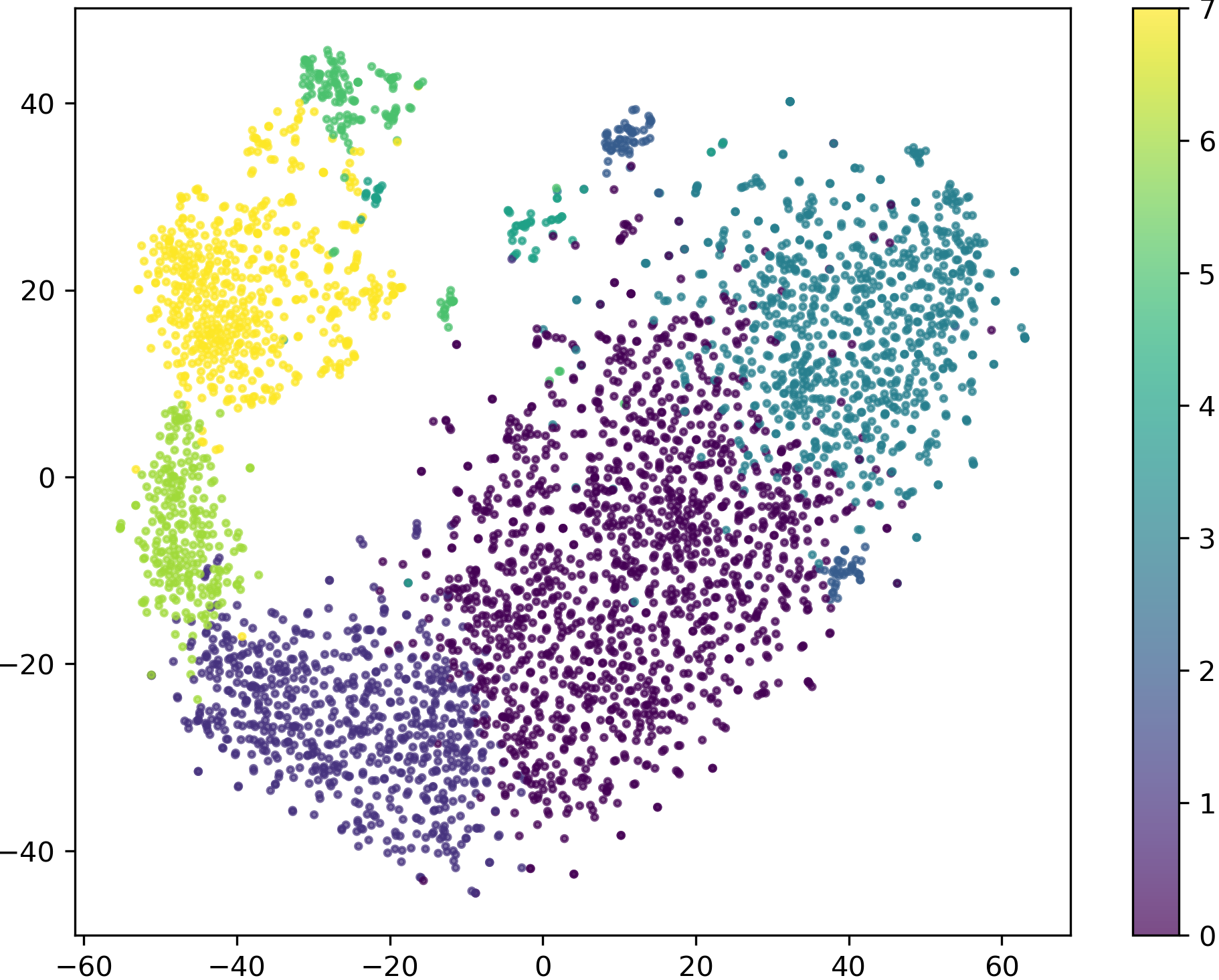}
        \caption{\themodelg}
        \label{fig:tsne_edge_louvain}
    \end{subfigure}
    \caption{Comparison of t-SNE visualizations on protein-level representations obtained from \themodel and \themodelg with the corresponding community labels detected by the Louvain algorithm on the PPI graph. Embeddings from \themodelg recapitulate the topological information.}
    \label{fig:tsne_louvain}
    \vspace{-5pt}
\end{wrapfigure}

\subsection{\textbf{(RQ5)} \themodelg Encodes Graph Relational Information}

To evaluate if \themodelg encodes graph relational information, we first conduct the graph-contextual protein language modeling on the PPI graph provided by \texttt{ogbn-proteins} dataset \cite{hu2020open}, which contains proteins from 8 species, to get a well-trained \themodelg. After training, we sample the same number of proteins from each species and obtain their representations using both \themodel and \themodelg. Since we sampled a subset of proteins (nodes) from the dataset, we can use them to construct a subgraph as well. We then use the Louvain algorithm \citep{de2011generalized} to detect 8 communities (corresponding to 8 species) in this subgraph. Next, we use t-SNE \citep{van2008visualizing} to reduce the dimensionality of both sets of protein embeddings obtained from \themodel and \themodelg and label each data point using its community membership. As shown in \Cref{fig:tsne_louvain}, the embeddings from \themodelg captures the graph topology much better than \themodel, which aligns with the community detection results. This suggests that our proposed graph context-aware second-stage training captures the topological information in PPI graphs as expected.

\begin{table}[h!]
\centering
\resizebox{\textwidth}{!}{%
\begin{tabular}{l c c c c c}
\toprule
\makecell{\textbf{Model} \\ \textbf{(\#Tokens trained)}} & 
\makecell{\textbf{PPI} \\ \textbf{graph}} & 
\makecell{\textbf{Contact} \\ \textbf{Map}} & 
\makecell{\textbf{Remote} \\ \textbf{Homology}} & 
\makecell{\textbf{Secondary} \\ \textbf{Structure}} & 
\makecell{\textbf{Stability}} \\
\midrule
ESM-2-650M (100B) &  None & 44.05 & $26.57 \pm 0.49 $  & $79.86 \pm 0.09$ & $0.763 \pm 0.01$ \\
ESM-2-G-650M (100B) & ogbn-proteins & 32.35 & $25.60 \pm 0.77 $  & $79.76 \pm 0.24 $ & $0.750 \pm 0.02$ \\
ESM-2-G-650M (100B) & ogbl-ppa & 26.66 & $27.18 \pm 0.63 $  & $79.91 \pm 0.24 $ & $0.753 \pm 0.01$ \\
\midrule
\themodel-740M (100B) & None & 47.10 & $35.14 \pm 1.69 $  & \cellcolor{gray!15}${85.07 \pm 0.03} $ & $0.794 \pm 0.01$ \\
\themodelg-740M (100B) & ogbn-proteins & {47.15} & \cellcolor{gray!30}$\mathbf{35.74 \pm 0.93}$  & $85.02 \pm 0.11$ & $0.801 \pm 0.01$ \\
\themodelg-740M (100B) & ogbl-ppa & 47.23 & \cellcolor{gray!15}$35.60 \pm 1.45$  & $85.01 \pm 0.03$ & $0.801 \pm 0.01$ \\
\midrule
ProtMamba-public & None & 10.96 & $17.82 \pm 1.85$  & $68.43 \pm 0.06$ & $0.726 \pm 0.01$ \\
CARP-640M-public & None & 25.83 & $28.0 \pm 0.8$  & $83.0 \pm 0.1$ & $0.720 \pm 0.01$ \\
ESM-2-650M-public (1T)  & None & \cellcolor{gray!30}\textbf{66.85} & $33.43 \pm 0.35 $  & $84.30 \pm 0.15$ & \cellcolor{gray!15}$0.804 \pm 0.01$ \\
\themodel-740M (1T) & None & \cellcolor{gray!15}{61.93} & $33.24 \pm 0.21 $  & \cellcolor{gray!30}$\mathbf{88.19 \pm 0.03} $ & \cellcolor{gray!30}$\mathbf{0.840 \pm 0.02}$ \\
\bottomrule
\end{tabular}
}
\caption{Evaluation on TAPE tasks in zero-shot (contact map) and supervised fine-tuning (remote homology and secondary structure) settings. We report the \emph{Precision@2/L} for Contact Map prediction, \emph{top-1 accuracy} for the Remote Homology fold-level test set, \emph{accuracy} for the 3-class secondary structure prediction on the CB513 test set, and \emph{spearman's rho}  for stability prediction, respectively. Values for CARP are taken from \citet{yang2024convolutions}. We perform 3 runs using different seeds to report the mean and standard deviation.
}
\label{tab:tape}
\end{table}

\subsection{\textbf{(RQ6)} Interaction Information Helps Downstream Tasks}

\paragraph{TAPE} Here we ask whether the graph relational information is helpful for common downstream tasks. We first use the remote homology detection and secondary structure prediction tasks from \texttt{TAPE} \citep{rao2019evaluating}, which represent protein-level and residue-level tasks, respectively. Our results in \Cref{tab:tape} show that \themodel achieves significantly better performance in both tasks compared to ESM-2 pretrained with the same number of tokens. Remarkably, \themodel and \themodelg even outperformed the public ESM-2 pretrained with 1T tokens, underscoring the sample efficiency of \themodels-based model architecture can translate to downstream tasks in a supervised fine-tuning setting. The marginal improvement of \themodelg over \themodel in remote homology tasks also suggests the information from the PPI graph helps determine protein's remote homologs, while not helpful in predicting their secondary structures at the AA level.  

\paragraph{ProteinGym} Next, we evaluate the predicted fitness landscape on zero-shot mutation effect prediction. It has been shown that pretrained pLMs can capture the fitness landscape of proteins without any further training \citep{meier2021language}. We use 217 deep mutational scan (DMS) datasets collected in \texttt{ProteinGym} \citep{notin2024proteingym}, which collectively measure the effects of 2.5 million substitution mutations to parent protein sequences. In \Cref{tab:proteingym}, we demonstrate \themodel achieved significantly better alignment with protein fitness compared to ESM-2 pretrained with the same number of tokens. Interestingly, we note that the graph-contextual training hurts the fitness landscapes of ESM-2 models, while \themodelg retained their zero-shot capabilities for protein fitness prediction. We hypothesize that the long graph context degrades the representation space of ESM-2, not for \themodelg. This highlights \themodels as a superior architectural design choice, demonstrating robustness in maintaining performance across various tasks while excelling in those that benefit from interaction information learned through graph-contextualized training, possibly due to its preferable context-length extrapolation property.

\begin{table}[h!]
\centering
\begin{adjustbox}{width=0.85\linewidth}
\begin{tabular}{l c c c}
\toprule
\textbf{Model (\#Tokens trained)} & \textbf{PPI graph} & \textbf{Spearman} & \textbf{NDCG} \\
\midrule
ESM-2-650M (100B) & None & $0.295 \pm 0.013$ & $0.695 \pm 0.008$ \\
ESM-2-G-650M (100B) & ogbn-proteins & $0.109 \pm 0.013$ & $0.642 \pm 0.008$ \\
ESM-2-G-650M (100B) & ogbl-ppa & $0.131 \pm 0.014$ & $0.644 \pm 0.007$ \\
\midrule
\themodel-740M (100B) & None & \cellcolor{gray!15}$0.378 \pm 0.008$ & \cellcolor{gray!30}$\mathbf{0.735 \pm 0.005}$ \\
\themodelg-740M (100B) & ogbn-proteins & \cellcolor{gray!30}$\mathbf{0.380 \pm 0.008}$ & \cellcolor{gray!15}$0.734 \pm 0.006$ \\
\themodelg-740M (100B) & ogbl-ppa & \cellcolor{gray!30}$\mathbf{0.380 \pm 0.008}$ & \cellcolor{gray!15}$0.734 \pm 0.006$ \\
\bottomrule
\end{tabular}
\end{adjustbox}
\caption{Evaluation on ProteinGym DMS substitutions benchmark. We report \emph{Spearman's correlation coefficient} and \emph{normalized discounted cumulative gain (NDCG)} between the log odds ratio and the experimentally measured protein fitness scores for each DMS assay.}
\label{tab:proteingym}
\end{table}

\subsection{(\textbf{RQ7}) Protein Function Prediction and Link Prediction on PPI Graph}
We evaluate \themodelg on two tasks: protein function prediction (\texttt{ogbn-proteins}) and PPI link prediction (\texttt{ogbl-ppa}). On \texttt{ogbn-proteins}, \themodelg achieves an \emph{accuracy} of 0.8925 $\pm$ 0.001, outperforming the state-of-the-art by 2.6\%. For \texttt{ogbl-ppa}, we leverage the learned embeddings from both \themodel and \themodelg to initialize the node attributes for GCN and GraphSAGE, evaluating these model variants. The results confirm that the embeddings improve performance. We conduct similar experiments on \texttt{ogbn-proteins}, further validating the effectiveness of capturing graph-contextual information. Additional details are provided in \Cref{app:graphdownstream}.

\section{Conclusion, Discussion, and Future Work} \label{sec:con}

In this work, we explored \themodel and \themodelg based on \themodels. We demonstrate \themodel's favorable neural scaling laws and length extrapolation property than Transformer-based ESM-2. We found that the length extrapolation property can facilitate the 3-D structure prediction of longer proteins and protein complexes. Specifically, \themodel outperformed ESM-2 by up to 30\% and 16\% on various downstream tasks when trained with 100B and 1T tokens, respectively. We also found that after training within graph context using random walk sampling, \themodelg can capture relational structure encoded in protein-protein interactions and improve remote homology prediction by more than 35\% compared to ESM-2. 

\themodel not only demonstrates superior performance on longer protein sequences but also outperforms ESM-2 on shorter protein sequences, highlighting a significant performance gap between SSMs and Transformers in protein language modeling. We hypothesize that this advantage may be attributed to the relatively small vocabulary size of protein sequences ($\sim$ 20 tokens), which allows SSMs to more effectively learn compressed state representations compared to natural languages, where the vocabulary size typically exceeds 50k tokens.

It is also possible that ESM-2 could incorporate more advanced architectural designs within its Transformer-based framework to narrow the performance gap, given the rapid advancements in text modeling and Transformer architectures. There are also emerging techniques that may further assist ESM-2 in bridging the gap in long-context modeling capability compared to \themodel.

In future work, we aim to (i) explore hybrid architectures that integrate multi-head attention with SSMs, (ii) investigate more advanced self-supervised training strategies to enhance the incorporation of graph-contextual information during the later stages of pre-training, e.g., contrastive learning using \emph{(positive, negative)} pairs of random walk paths to reinforce locality relationships, and (iii) develop more principled approaches for negative sampling to better contrast with \emph{positive} random walk paths. Some other directions include (i) approximate permutation-invariant graph context learning for pLMs, e.g. using permutation group \citep{huang2022going}, (ii) explore other graph context extraction methods instead of random walk, e.g. graph skeleton tree \citep{huang2023growing}.

We also think it deserves to expand the application scope of \themodel across several key areas: (i) understanding viral protein sequences, which are characterized by extended sequence lengths; (ii) enhancement of protein co-regulation and functional prediction capabilities \citep{hwang2024genomic}; and (iii) advanced protein design tasks requiring expanded contextual understanding, such as protein inpainting. We anticipate that \themodel's capabilities will enable novel applications beyond these identified domains, presenting significant opportunities for further exploration in the field of protein modeling and design.



\bibliography{iclr2025_conference}
\bibliographystyle{tmlr}

\newpage

\appendix

\setcounter{tocdepth}{2}
\tableofcontents
\allowdisplaybreaks
\begin{appendices}

\section{Definitions of Criterions Used in \Cref{tab:sum}}\label{app:cri}

Here, we provide clear definitions for all criteria listed in \Cref{tab:sum} and explain the rationale for whether each method possesses the corresponding feature.
\begin{itemize}
    \item \textbf{Universality}: We define this criteria and justify whether a method possesses this feature (1) if it is \textit{learning universal, cross-family representations of protein space} as defined in previous works of protein representation learning~\citep{alley2019unified, detlefsen2022learning}; and (2) if it is pretrained on universal protein dataset, e.g. the \textit{Universal} Protein Reference Clusters (UniRef) dataset which contains universal protein sequence resources; (3) if the learned expressive protein/AA embeddings can achieve decent performance across a variety of downstream tasks, which demonstrates the universality of such representations.
    \item \textbf{Fine granularity}: given that the protein sequence is composed of high-resolution AA tokens, we define this criteria according to whether the model can take in AA token as input. Note that Graph-Mamba~\citep{wang2024graphmambalongrangegraphsequence} and GMN~\citep{behrouz2024graphmambalearninggraphs} can only handle inputs like PPI graphs where each node corresponds to an individual protein.
    \item \textbf{Long-context capability (Handleability)}: we define this criteria according to whether the model can take in extremely long (e.g. $> 16K$ sequence length) input sequence. Note that Transformers naturally lack this feature and ProtMamba~\citep{sgarbossa2024protmamba} discards this capability by improperly using positional encodings (PEs)\footnote{\url{https://github.com/Bitbol-Lab/ProtMamba-ssm/blob/8befff756b2db7b6dc56d0a07163eb02e27b2731/ProtMamba_ssm/modules.py}} which makes it unable to extrapolate to sequences $> 2048$ on any downstream tasks that need fine-tuning on longer sequences as we discussed in the main text. ProtMamba used learnable PE as an improper design choice that will do harm to the length extrapolation capability of the model since such PE has been demonstrated as an un-extrapolatable PE in many works~\citep{zhao2023length,sun2022length}.
    \item \textbf{Long-context capability (Performance)}: we define this criteria according to whether the model can perform well across a variety of downstream tasks that need long-context dependencies (e.g. structure prediction on protein complex, remote homology prediction).
    \item \textbf{Graph context}: we define this criteria according to whether the model can take into account the protein interaction graph contexts. All models purely based on protein sequence fail.
    \item \textbf{Large-scale model}: we define this criteria according to whether the model has been scaled up to a fairly large size (e.g. the number of trainable model parameters closer to 1B).
\end{itemize}

\section{More Discussion on General pLMs} \label{app:plm}
As noted earlier, protein sequences, represented as strings of amino acid letters, are well-suited to LMs that can capture complex dependencies among amino acids~\citep{OFER20211750}. pLMs~\citep{hu2022protein} have emerged as promising tools for learning protein sequences. This section introduces LSTM-based pLMs, followed by Transformer-based pLMs, detailing their implementation strategies and applications, particularly for protein structure prediction.

\cite{MichaelSchantzKlausen2018NetSurfP20IP} developed a combination of convolutional and LSTM neural networks to predict various protein structural features, such as solvent accessibility, secondary structure, structural disorder, and torsion angles ($\varphi$, $\psi$) for each residue. Models like SPIDER3-Single~\citep{RhysHeffernan2018SinglesequencebasedPO} focus on single sequences rather than relying on multiple sequence alignments (MSAs). Similarly, models such as DeepPrime2Sec~\citep{EhsaneddinAsgari2019DeepPrime2SecDL} and SPOT-1D-Single~\citep{JaspreetSingh2021SPOT1DSingleIT} share comparable training objectives and architectures. Furthermore, models like DeepBLAST~\citep{JamesTMorton2020ProteinSA}, SPOT-1D-LM~\citep{JaspreetSingh2021SPOT1DLMRA}, and SPOT-Contact-Single~\citep{JaspreetSingh2021SPOTContactSingleIS} utilize embeddings from pre-trained pLMs for downstream tasks such as contact map and function prediction.

However, the TAPE benchmark~\citep{rao2019evaluating} highlighted opportunities for innovative design and training methods beyond traditional LSTMs and Transformers. UniRep~\citep{EthanCAlley2019UnifiedRP}, for example, employs a multiplicative LSTM (mLSTM)\citep{BenKrause2016MultiplicativeLF} to condense arbitrary protein sequences into fixed-length vectors, capturing long-range dependencies. Similarly, UDSMProt\citep{Strodthoff:2019universal} and SeqVec~\citep{MichaelHeinzinger2019ModelingTL} utilize LSTM variants to develop rich, transferable representations. ProSE~\citep{TristanBepler2021LearningTP} enhances these representations with structural supervision through residue-residue contact loss and structural similarity prediction, while CPCProt~\citep{AmyXLu2020SelfSupervisedCL} leverages InfoNCE loss to maximize mutual information in protein embeddings.

ProtTrans~\citep{AhmedElnaggar2021ProtTransTC} trained extensive models (including T5, ELECTRA, ALBERT, XLNet, BERT, and Transformer-XL) on sequences comprising 393 billion amino acids across 5616 GPUs and one TPU Pod. ESM-1b~\citep{AlexanderRives2019BiologicalSA} demonstrates how deep Transformers, coupled with a masking strategy, can build intricate context-aware representations. The results from ProtTrans and ESM-1b suggest that large-scale pLMs can effectively learn the grammar of proteins, even without evolutionary data. Furthermore, PMLM~\citep{LiangHe2022PretrainingCP} enhances model performance on the TAPE contact benchmark by accounting for dependencies among masked tokens, indicative of inter-residue coevolution.

Incorporating additional data such as MSAs, functions, structures, and biological priors can enrich protein embeddings. For instance, the MSA Transformer~\citep{RoshanRao2021MSAT} adapts Transformer LMs to handle sets of sequences, utilizing alternating attention mechanisms. ProteinBERT~\citep{DanOfer2021ProteinBERTAU} integrates sequence information with Gene Ontology (GO) annotations to predict diverse protein functions, while OntoProtein~\citep{zhang2022ontoprotein} leverages GO as a factual knowledge graph. The PEER benchmarks~\citep{MinghaoXu2022PEERAC} demonstrate the importance of selecting suitable auxiliary tasks to enhance model performance across a variety of protein-related tasks.

\paragraph{Protein Structure Prediction}
Early pLMs~\citep{MichaelSchantzKlausen2018NetSurfP20IP, RhysHeffernan2018SinglesequencebasedPO, EhsaneddinAsgari2019DeepPrime2SecDL} primarily predicted structural features, which are essential for constructing 3D protein structures. Recent models aim to predict protein structures end-to-end. Evoformer, a core module in the AF2 network~\citep{jumper2021highly}, exemplifies this with its sophisticated design that includes axial attention and updates to pair representations ensuring consistency principles like the triangle inequality.

\paragraph{Other Applications}
ProGen~\citep{madani2020progen} exemplifies models trained on sequences conditioned on specific protein properties. In contrast, newer models like ProGen2~\citep{ErikNijkamp2022ProGen2ET} and AminoBERT~\citep{RatulChowdhury2021SinglesequencePS} illustrate the expansion of pLM applications to include tasks like antibody structure prediction, demonstrating the versatile utility of pLMs across a range of biological research and clinical applications.

\section{Datasets, Tasks, and Metrics} \label{app:data}

\subsection{Protein Sequence Datasets}

We first describe the \texttt{Unified Reference Protein (UniRef)} dataset\citep{suzek2015uniref}, which provides clustered sets of protein sequences from the UniProt Knowledgebase (UniProtKB) \citep{boutet2016uniprotkb} and selected UniParc records. It's designed to speed up protein sequence analysis by reducing the redundancy of sequences at different levels without losing the coverage of sequence space. Here are the key features of the \texttt{UniRef} dataset:
\begin{itemize}
    \item \texttt{UniRef100:} This dataset includes all the protein sequences from UniProtKB and selected UniParc records, clustered by exact sequence matches. It provides comprehensive coverage and serves as the basis for the other two datasets.
    \item \texttt{UniRef90:} This set clusters sequences that have at least 90\% sequence identity and 80\% overlap in alignment, compressing the dataset while still preserving most of the sequence diversity. It is used for high-throughput and large-scale analysis where a balance between speed and coverage is needed.
    \item \texttt{UniRef50:} This dataset clusters sequences with at least 50\% sequence identity and 80\% overlap in alignment, further reducing the dataset size and redundancy. It's intended for rapid scans and for exploring broad phylogenetic relationships.
\end{itemize}

Each entry in a UniRef dataset represents a cluster and contains the sequence of the representative protein (the longest sequence or the one with the most annotations), along with a list of all the cluster members. These datasets are useful for various bioinformatics tasks such as sequence alignment, phylogenetic analysis, and functional annotation, as they allow researchers to handle large volumes of sequence data more efficiently. 

We use the 2024-01 release of UniRef\footnote{\url{https://ftp.uniprot.org/pub/databases/uniprot/previous_releases/release-2024_01/uniref/}}, and preprocessed by removing de-novo designed proteins by filtering out protein sequences annotated by \texttt{Tax=synthetic construct}. Next, we randomly sample 250,000 sequences from \texttt{UniRef90} as the validation set to report evaluation losses for pretraining protein language models (pLMs). To remove sequences from training sets (\texttt{UniRef50} and \texttt{UniRef90}) that are highly similar to the validation set, we use the training sets as query databases and validation set as a target database by mmseqs2 \citep{steinegger2017mmseqs2} with the following command: \texttt{mmseqs search --min-seq-id 0.5 --alignment-mode 3 --max-seqs 300 -s 7 -c 0.8 --cov-mode 0}. 

\begin{figure}
    \centering
    \includegraphics[width=\linewidth]{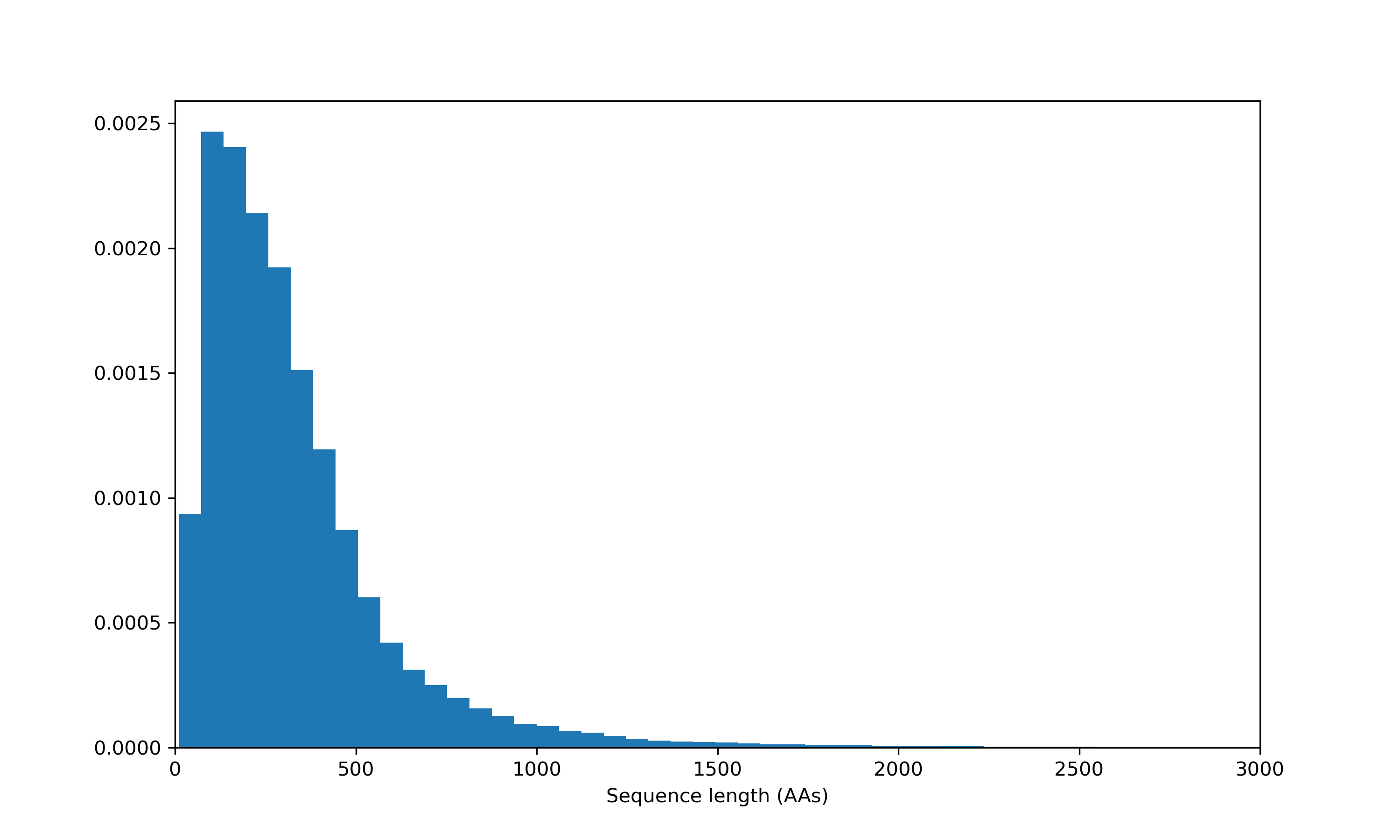}
    \caption{The sequence length distribution of \texttt{UniRef90}.}
    \label{fig:ur90_hist}
\end{figure}

For the first-stage training of \themodel and ESM-2 models, we use the \texttt{UniRef50} training set; and for the scaling law and length extrapolation experiments, we use the \texttt{UniRef90} set. We also provide the histogram of \texttt{UniRef90} in terms of sequence length in \Cref{fig:ur90_hist}. The training and evaluation sets are randomly sampled from the entire set, which should follow the same distribution. The average lengths of \texttt{UniRef90} and \texttt{UniRef50} are shown in \Cref{tab:ur_length}

\begin{table}[ht]
    \centering
    \begin{tabular}{lccc}
        \toprule
        \textbf{Database} & \textbf{Tokens} & \textbf{Num sequences} & \textbf{Average length} \\
        \midrule
        UniParc   & 221.7B & 577.8M & 383.6bp \\
        UniRef100 & 144.3B & 376.6M & 383.2bp \\
        {UniRef90}  & {61.2B} & {179.5M} & {341.2bp} \\
        UniRef50  & 17.8B  & 62.8M  & 284.4bp \\
        \bottomrule
    \end{tabular}
    \caption{Average length of \texttt{UniRef}.}
    \label{tab:ur_length}
\end{table}

\subsection{Structure Prediction Datasets}

\texttt{CASP15-multimers} is a subset derived from the 15th edition of Critical Assessment of Protein Structure Prediction (CASP) challenge \citep{kryshtafovych2023critical}, specifically focusing on predicting the structures of protein complexes or multimers. In \texttt{CASP15}, the multimer track evaluates the ability of protein structure prediction methods to accurately model the quaternary structures of protein assemblies. \texttt{CASP15-multimers} includes 52 protein complexes.

\texttt{CASP14} is a dataset from the 14th edition of the CASP challenge \citep{kryshtafovych2021critical}. CASP is a biennial experiment that assesses the state-of-the-art methods for protein structure prediction. \texttt{CASP14} covers a wide range of protein structure prediction tasks, including free modeling (FM), template-based modeling (TBM), and the prediction of protein domains with challenging folds. \texttt{CASP14} includes 37 protein structures.

\texttt{Benchmark2} \citep{ghani2021improved} is a dataset commonly used in the field of protein-protein docking and structure prediction. It is a curated collection of protein complexes that have been used extensively to evaluate the performance of computational methods in predicting the binding orientation of protein complexes. The dataset includes both bound and unbound forms of protein structures, providing a challenging testbed for docking algorithms. \texttt{Benchmark2} includes 17 heterodimers structures.

\paragraph{Metrics}
(i) \texttt{Frame Aligned Point Error (FAPE)} measures the error in aligning one set of points (predicted) to another (target), considering both translational and rotational components. The error is calculated on a per-point basis between corresponding points in the two sets, after aligning them using a frame of reference.
The error is normalized by the size of the objects involved, which allows it to be invariant to the absolute size of the objects, making it suitable for tasks involving objects of varying scales.
The loss is usually computed as an L1 or L2 norm of the differences in the aligned coordinates, which provides a straightforward gradient for optimization.
(ii) \texttt{Distogram Loss} aims to improve the separability of feature distributions for different classes. It works by encouraging the histograms (or distributions) of distances within classes (positive pairs) to be distinct from histograms of distances between classes (negative pairs).
The loss function is designed to increase the overlap between histograms of positive pairs while decreasing the overlap for negative pairs. This is achieved by calculating the probability of a randomly chosen positive pair having a smaller distance than a randomly chosen negative pair.
Typically, a differentiable approximation of the histogram is used, and the optimization focuses on adjusting the model parameters to achieve the desired separation in the histograms' distributions.
(iii) \texttt{pTM score} is used to assess the structural similarity between two proteins, normalized for protein size. TM scores range from 0 to 1, where a score higher than 0.5 generally indicates a model of correct topology and a score below 0.17 suggests random similarities. TM-score is more sensitive to the global fold of a protein than to the specific atomic positions, making it particularly useful in assessing larger, domain-level accuracies. (iv) \texttt{LDDT} is a local superposition-free score that evaluates the local accuracy of a protein model by comparing distances between all atom pairs within defined cutoffs in both the predicted and reference models. It can provide a more detailed view of the quality of a pLM at the residue level.

\subsection{TAPE and ProteinGym}
\paragraph{Tasks Assessing Protein Embeddings (TAPE)} TAPE \citep{rao2019evaluating} is a set of five biologically relevant semi-supervised learning tasks spread across different domains of protein biology. We adopted four tasks: 1) Secondary Structure prediction, and 2) Remote Homology Detection, 3) Stability prediction, 4) Fluorescence prediction. Secondary structure prediction is a sequence-to-sequence task where each input amino acid is mapped to one of the three labels from $\{\text{Helix}, \text{Strand}, \text{Other} \}$. It probes the model's ability to learn local structure. Remote Homology Detection is a sequence classification task where each input protein is mapped to a label $\{1,\cdots,1195 \}$, representing different possible protein folds. This task measures models' capability to detect structural similarity across distantly related proteins. Stability and Fluorescence predictions are sequence-level regression tasks that probe the model's ability to predict the mutant sequences' properties, which are the thermostability and the flurescent intensity, respectively.

\paragraph{ProteinGym}  ProteinGym \citep{notin2024proteingym} is collection of benchmarks aiming at comparing the ability of models to predict the effects of protein mutations. We use the DMS Substitution subset from ProteinGym, which covers 2.5 million mutants from across 217 assays.    

\subsection{Datasets for Protein Function Prediction and Link Prediction on PPI Graph} \label{app:graphtasks}
\paragraph{ogbn-proteins}
The \texttt{ogbn-proteins} dataset is structured as an undirected, weighted graph with nodes and edges categorized by species. Nodes in this graph represent proteins, while the edges denote various biologically significant associations such as physical interactions, co-expression, or homology \citep{szklarczyk2019string}. Each edge is associated with an 8-dimensional feature vector, where each dimension quantifies the confidence level of a particular association type on a scale from 0 to 1—with higher values indicating greater confidence. The dataset includes proteins from eight different species.
The objective is to predict protein functions using a multi-label binary classification approach, where there are 112 different functions to predict. The performance metric used is the average ROC-AUC score across these 112 prediction tasks.
The dataset is divided into training, validation, and test sets based on the species of the proteins. This split strategy is designed to assess how well models can generalize across different species.
Notably, the \texttt{ogbn-proteins} dataset lacks specific input features for nodes but includes features for over 30 million edges. In baseline experiments, a simple approach is taken where the node features are derived by averaging the features of incoming edges to each node.

\paragraph{ogbl-ppa}
The \texttt{ogbl-ppa} dataset is an undirected, unweighted graph where nodes represent proteins from 58 different species, and edges illustrate biologically significant relationships between proteins, including physical interactions, co-expression, homology, or genomic neighborhood \citep{szklarczyk2019string}. Each node is described by a 58-dimensional one-hot vector indicating the species of the protein.
The objective is to predict potential new association edges based on the provided training edges. The model's performance is evaluated on its ability to prioritize positive test edges over negative ones. Each positive edge in the validation or test set is ranked against 3,000,000 randomly selected negative edges. The effectiveness of the model is measured using the Hits@K metric, where K = 100 is determined to be an effective threshold in initial experiments. This metric, which requires the model to consistently rank positive edges above a vast majority of negative edges, poses a greater challenge than the ROC-AUC metric.
The dataset divides edges into training, validation, and test sets based on the method used to determine the associations. Training edges consist of associations identified either through high-throughput methods (such as automated, large-scale experiments) or computationally (e.g., through text mining). Conversely, validation and test edges are derived from protein associations verified through low-throughput, labor-intensive experiments in the lab \citep{macarron2011impact,bajorath2002integration,younger2017high}. The primary challenge is to predict specific types of protein associations, like physical protein-protein interactions, based on other more readily measurable types of associations that are correlated with the target interactions.

\begin{table}[h!]
\centering
\begin{tabular}{ccccccc}
\toprule
\textbf{Name} & \textbf{\#Nodes}  & \textbf{\#Edges}   & \textbf{Split} & \textbf{Task} & \textbf{Metric}   \\
\midrule
ogbn-proteins & 132,534  & 39,561,252      & Species    & Binary classification & ROC-AUC  \\
ogbl-ppa      & 576,289  & 30,326,273    & Throughput & Link prediction       & Hits@100 \\
\bottomrule
\end{tabular}
\caption{Summary of OGB datasets.}
\label{tab:dataset_summary}
\end{table}

\section{Experimental Setup}\label{app:exp}

\subsection{Hardware and Software}
All experiments are run on NVIDIA A100 Tensor Core GPU except \texttt{ogbn-proteins} and \texttt{ogbl-ppa}, which are run on NVIDIA A10G Tensor Core GPUs. For core software packages in main experiments, we use Python 3.10, PyTorch 2.1.0, Transformers 4.41.2, DeepSpeed 0.14.4, Accelerate 0.27.2, mamba-ssm 2.2.0, datasets 2.20.0, Triton 2.0.0, and CUDA Toolkit 12.1. For some downstream tasks, the dependencies and package version will be adjusted accordingly. For \texttt{ogbn-proteins} and \texttt{ogbl-ppa}, we add several new packages: PyTorch Geometric 2.5.3, torch-cluster 1.6.3, torch-scatter 2.1.2, torch-sparse 0.6.18, torch-spline-conv 1.2.2, and OGB 1.3.6.

\subsection{Masked Language Modeling}
The input to the model consists of raw protein sequences, which are tokenized into individual amino acids. A subset of these amino acids is randomly selected and masked during training. Typically, 15\% of the amino acids in a sequence are selected for masking. Of these selected tokens, 80\% are replaced with a special \texttt{[MASK]} token, 10\% are replaced with a random amino acid, and the remaining 10\% are left unchanged. The model is trained to predict the identity of these masked amino acids using a cross-entropy loss.

The training process is conducted using the AdamW optimizer with a learning rate that is linearly warmed up for a small percentage of the total training steps, followed by a cosine decay schedule. The batch size and learning rate are chosen based on the model size and computational resources, with a total number of tokens approximately equal to 0.5M\footnote{For 100M, 340M, 740M, and 1.3B of \themodel, we train on 16, 32, 64, 128 A100s, respectively. The local batch size and the gradient accumulation steps are dynamically adjusted to ensure the model fits in.} and learning rates are set as $2 \times 10^{-4}$. Gradient clipping is often applied to stabilize the training. Additionally, a 0.1 weight decay is applied. 

For second-phase training with graph context, we sample random walks with the context length $l=5$ on the given PPI graph (\texttt{ogbn-proteins} or \texttt{ogbl-ppa}) and retrieve the corresponding protein sequence for each node from String database \citep{szklarczyk2019string}. We then add 4 special tokens to indicate the begin and end of the node, the edge and non-edge in the random walk path. We continue MLM training the 740M \themodel on 20B more tokens on random walks to get \themodelg, during which we freeze the parameters in SSMs and linear projection layers except the input \& output embeddings and normalization layers.
We summarize the key hyperparameters in \cref{tab:mlm_hp}.

\subsection{Structure Prediction with LMFold}
To train LMFold, we use the FAPE and distogram losses introduced in AlphaFold2 \citep{jumper2021highly}, as well as heads for predicting LDDT and the pTM score. We weigh these 4 loss terms using the default constants proposed in OpenFold \citep{ahdritz2024openfold}.
We used Adam with $\beta_1= 0.9$, $\beta_2= 0.99$, and $\epsilon=1^{-6}$ as optimizers, and warmed up the learning rate linearly over the first 1,000 iterations from 0 to $1^{-3}$. We use a per-GPU batch size of 1 training on 32 NVIDIA A100 GPUs, leading to a global batch size of 32. 

\begin{table}[h!]
\centering
\begin{tabular}{l c c c c}
\toprule
\textbf{Hyperparameters} & \textbf{100M} & \textbf{340M} & \textbf{740M} & \textbf{1.3B} \\
\midrule
Peak learning rate & \multicolumn{4}{c}{$2 \times 10^{-4}$} \\
Global batch size & \multicolumn{4}{c}{0.5M tokens} \\
Block size & \multicolumn{4}{c}{1024} \\
Warm-up steps & \multicolumn{4}{c}{2000} \\
Adam betas & \multicolumn{4}{c}{$\beta_1=0.9, \beta_2=0.95$} \\
Maximum gradient norm & \multicolumn{4}{c}{0.5} \\
Precision & \multicolumn{4}{c}{BF16} \\
Optimizer & \multicolumn{4}{c}{AdamW} \\
Learning rate scheduler & \multicolumn{4}{c}{cosine} \\
Weight decay & \multicolumn{4}{c}{0.1} \\
Length of random walks & \multicolumn{4}{c}{5} \\
\midrule
Hidden size & 768 & 1024 & 1536 & 2048 \\
Number of \themodels blocks & 24 & 48 & 48 & 48 \\
\bottomrule
\end{tabular}
\caption{Summary of MLM training hyperparameters for \themodel and \themodelg.}
\label{tab:mlm_hp}
\end{table}

\begin{table}[h!]
\centering
\begin{tabular}{l c c c c}
\toprule
\textbf{Hyperparameters} & \textbf{150M} & \textbf{300M} & \textbf{650M} & \textbf{1.0B} \\
\midrule
Peak learning rate & \multicolumn{4}{c}{$2 \times 10^{-4}$} \\
Global batch size & \multicolumn{4}{c}{0.5M tokens} \\
Warm-up steps & \multicolumn{4}{c}{2000} \\
Adam betas & \multicolumn{4}{c}{$\beta_1=0.9, \beta_2=0.98$} \\
Maximum gradient norm & \multicolumn{4}{c}{1.0} \\
Precision & \multicolumn{4}{c}{BF16} \\
Optimizer & \multicolumn{4}{c}{AdamW} \\
Learning rate scheduler & \multicolumn{4}{c}{cosine} \\
Weight decay & \multicolumn{4}{c}{0.01} \\
Length of random walks & \multicolumn{4}{c}{5} \\
\midrule
Hidden size & 640 & 960 & 1280 & 1280 \\
Intermediate size & 2560 & 3840 & 5120 & 7680 \\
Number of hidden layers & 30 & 30 & 33 & 33 \\
\bottomrule
\end{tabular}
\caption{Summary of MLM training hyperparameters for ESM-2.}
\label{tab:esm2_mlm_hp}
\end{table}

\subsection{Evaluation on TAPE tasks}

To evaluate pretrained pLMs on TAPE Remote Homology prediction and Secondary Structure prediction, we followed the evaluation setting from \cite{rao2019evaluating}. For the Remote Homology task, we add a two-layered MLP (with 512 as the intermediate dimension) on top of the protein-level embeddings from a pLM. The protein-level embeddings are calculated as the average of token-level embeddings. For the Secondary Structure prediction task, we used a single-layered MLP taking the token-level embeddings from the LM directly to make a token-level 3-class classification.  

Then, we fine-tune the parameters of the LM and the MLP prediction head end-to-end on the training set and perform early stopping on the validation set. We report the top-1 accuracy on the fold-level hold-out test set for Remote Homology task, accuracy on the CB513 test set for the Secondary Structure prediction task, and spearman's rho on Stability prediction task, respectively. The detailed hyperparameters we used are listed in \Cref{tab:tape_hp}.

\begin{table}[h!]
\centering
\begin{tabular}{l c c c}
\toprule
\textbf{Hyperparameters} & \textbf{Remote Homology} & \textbf{Secondary Structure} & \textbf{Stability} \\
\midrule
Batch size &  \multicolumn{2}{c}{16} & 512  \\
Number of warm-up steps &  \multicolumn{3}{c}{5000} \\
Early stopping patience &  \multicolumn{3}{c}{25 epochs} \\
Max number of epochs &  \multicolumn{3}{c}{100} \\
Learning rate decay schedule &  \multicolumn{3}{c}{cosine} \\
Optimizer&  \multicolumn{3}{c}{AdamW} \\
Adam betas&  \multicolumn{3}{c}{$\beta_1=0.9, \beta_2=0.98$} \\
Peak learning rate & 1e-5 & 5e-5 & 1e-4  \\
Prediction head &  2-layered MLP & 1-layered MLP & 2-layered MLP \\
\bottomrule
\end{tabular}
\caption{Hyperparameters used for fine-tuning protein language models for TAPE tasks.
}
\label{tab:tape_hp}
\end{table}

\subsection{Evaluating on ProteinGym}
To evaluate pretrained pLMs on the ProteinGym DMS Substitution benchmarks, we adopt the masked-marginals heuristic \citep{meier2021language} to predict protein fitness in a zero-shot setting. The masked-marginals method scores mutations (mt) using the log odds ratio at the mutated position over wild-type (wt), assuming an additive model when multiple mutations $T$ exist in the same protein sequence: 
$$\sum_{t\in T}\log{p(x_t = x_t^{mt} | x_{\backslash T}}) - \log{p(x_t = x_t^{wt} | x_{\backslash T}})$$ 
We then compute Spearman's correlation coefficient and normalized discounted cumulative gain (NDCG) between the log odds ratio and the experimentally measured protein fitness scores for each DMS assay. The Spearman's correlation coefficients are aggregated across 217 DMS assays using the provided code.

\subsection{Protein Function Prediction and PPI Link Prediction}
For protein function prediction on \texttt{ogbn-proteins}, we use GIPA \citep{li2023gipa} as the graph neural network (GNN) backbone with our learned protein sequence embeddings as the node attributes initialization. We report the hyperparameters (HPs) we use to train the model in \Cref{tab:hp_gipa}. Furthermore, we conduct more ablation studies on the GNN backbone to test if the contribution of learned protein embeddings is independent of the architecture design choice. We choose two popular GNNs, i.e. GCN \citep{kipf2016semi} and GraphSAGE \citep{hamilton2017inductive} to evaluate the embeddings. The HPs are shown in \Cref{tab:hp_gcn_graphsage}.

For PPI link prediction on \texttt{ogbl-ppa}, we use a combined backbone of NGNN \citep{song2021network} and SEAL \citep{zhang2018link} with labeling tricks \citep{zhang2021labeling}. We summarize the HPs used in this backbone in \Cref{tab:hp_ngnn_seal}. Note that there is a ratio $k$ used in SortPooling \citep{zhang2018end}. We also perform similar ablation studies on this task using GCN and GraphSAGE with the same set of HPs reported in \Cref{tab:hp_gcn_graphsage}.

\begin{table}[h!]
\centering
\begin{minipage}{0.32\textwidth}
    \centering
    \begin{tabular}{l c}
    \toprule
    \textbf{Hyperparameters} & \textbf{Value} \\
    \midrule
    \# of epochs & 1500 \\
    \# of heads & 20 \\
    Learning rate & 0.01 \\
    \# of layers & 6 \\
    \# of hidden units & 50 \\
    Dropout rate & 0.4 \\
    \bottomrule
    \end{tabular}
    \caption{HPs of GIPA backbone.}
    \label{tab:hp_gipa}
\end{minipage}%
\hfill
\begin{minipage}{0.32\textwidth}
    \centering
    \begin{tabular}{l c}
    \toprule
    \textbf{Hyperparameters} & \textbf{Value} \\
    \midrule
    \# of epochs & 1000 \\
    \# of heads & 20 \\
    Learning rate & 0.01 \\
    \# of layers & 3 \\
    \# of hidden units & 256 \\
    Dropout rate & 0.3 \\
    \bottomrule
    \end{tabular}
    \caption{HPs of GCN/SAGE.}
    \label{tab:hp_gcn_graphsage}
\end{minipage}%
\hfill
\begin{minipage}{0.32\textwidth}
    \centering
    \begin{tabular}{l c}
    \toprule
    \textbf{Hyperparameters} & \textbf{Value} \\
    \midrule
    \# of epochs & 50 \\
    $k$ of SortPooling & 0.6 \\
    Learning rate & 0.00015 \\
    \# of layers & 3 \\
    \# of hidden units & 48 \\
    Dropout rate & 0.0 \\
    \bottomrule
    \end{tabular}
    \caption{HPs of NGNN/SEAL.}
    \label{tab:hp_ngnn_seal}
\end{minipage}
\end{table}

\section{More results and details on TAPE}
For the Jacobian contact map prediction task we reported in the main paper, we adopted the methods from \cite{zhang2024protein} to use categorical Jacobian matrices computed from protein language models as the zero-shot prediction for protein contact maps and report the precision@2/L (L is the length of a protein sequence) on the validation set of ProteinNet dataset \citep{alquraishi2019proteinnet}.  
\begin{table}[h!]
\centering
\begin{tabular}{l c c}
\toprule
\textbf{Model (\#Tokens trained)} & \textbf{PPI graph} & \textbf{Fluorescence} \\
\midrule
ESM-2-650M (100B) & None & $0.695 \pm 0.002$ \\
ESM-2-G-650M (100B) & ogbn-proteins & $0.694 \pm 0.002$ \\
ESM-2-G-650M (100B) & ogbl-ppa & $0.693 \pm 0.001$ \\
\midrule
\themodel-740M (100B) & None & $0.692 \pm 0.002$ \\
\themodelg-740M (100B) & ogbn-proteins & $0.709 \pm 0.003$ \\
\themodelg-740M (100B) & ogbl-ppa & $0.693 \pm 0.002$ \\
\midrule
ProtMamba-public & None & $0.688 \pm 0.005$ \\
CARP-640M-public & None & $0.680 \pm 0.002$ \\
ESM-2-650M-public (1T) & None & $0.688 \pm 0.001$ \\
\themodelg-740M (100B) & ogbl-ppa & $0.691 \pm 0.003$ \\
\bottomrule
\end{tabular}
\caption{Evaluation on TAPE tasks in supervised fine-tuning setting. We report the Spearman’s correlation coefficients for the test sets for Fluorescence prediction tasks. We perform 3 runs using different seeds to report the mean and standard deviation.
}
\label{tab:tape_app}
\vspace{-15pt}
\end{table}

\section{More results on ProteinGym}
We report more results and comparisons on ProteinGym with some other baselines, i.e. PoET \citep{truong2023poet},
TranceptEVE-L \citep{notin2022trancepteve}, and SaProt \citep{su2023saprot}, which use additional information like structural tokens.

The results demonstrate that \themodel performs better with less training tokens, as shown in \Cref{tab:proteingym}; however, with sufficient training tokens, ESM2 can achieve comparable or even better performance on this task, as shown in \Cref{tab:proteingym_app}. The sequence-only methods like ESM2 and \themodel have lower numbers than the other baselines, showing that structure information is useful for ProteinGym DMS substitutions.

\begin{table}[h!]
\centering
\begin{adjustbox}{width=0.85\linewidth}
\begin{tabular}{l c c c}
\toprule
\textbf{Model (\#Tokens trained)} & \textbf{PPI graph} & \textbf{Spearman} & \textbf{NDCG} \\
\midrule
ESM-2-650M-public (1T) & None & $0.414 \pm 0.011$ & $0.747 \pm 0.005$ \\
ESM-2-3B-public (1T) & None & $0.406 \pm 0.011$ & $0.755 \pm 0.004$ \\
\themodel-740M (1T) & None & ${0.388 \pm 0.009}$ & $0.751 \pm 0.004$ \\
\hline
PoET \citep{truong2023poet} & None & $0.479$ & N/A \\
TranceptEVE-L \citep{notin2022trancepteve} & None & $0.454$ & $0.786$ \\
SaProt \citep{su2023saprot} & None & $0.457$ & $0.768$ \\
\bottomrule
\end{tabular}
\end{adjustbox}
\caption{Evaluation on ProteinGym DMS substitutions benchmark. We report \emph{Spearman's correlation coefficient} and \emph{normalized discounted cumulative gain (NDCG)} between the log odds ratio and the experimentally measured protein fitness scores for each DMS assay.}
\label{tab:proteingym_app}
\end{table}

\section{Robustness of downsampling for structure prediction training set}
We perform three 1.5\% downsampling using three random seeds and retrain both our LC-PLM and ESM-2. We find that the downsampling is very robust to the performance with a small standard deviation that is close to (even smaller than) the standard deviation of using the same train set but training with different random seeds as we reported in \Cref{tab:sp_app}.
\begin{table}[h!]
\centering
\begin{adjustbox}{width=0.99\linewidth, center}
\begin{tabular}{l c c c}
\toprule
\textbf{Model (\#Tokens trained)} & \textbf{CASP15-multimers} & \textbf{CASP14} & \textbf{Benchmark2} \\
\midrule
ESM-2-650M (100B)    & $0.4132 \pm 0.0065$  & $0.3437 \pm 0.0039$  & $0.4773 \pm 0.0092$  \\
\themodel-740M (100B)  & ${0.5004 \pm 0.0139}$  & ${0.4244 \pm 0.0053}$  & ${0.6290 \pm 0.0121}$  \\
\midrule
ESM-2-650M-public (1T)  & ${0.5128 \pm 0.0003}$  & ${0.4421 \pm 0.0023}$  & ${0.6844 \pm 0.0059}$  \\
\bottomrule
\end{tabular}
\end{adjustbox}
\caption{Structure prediction performance (\emph{TM score}) on \texttt{CASP15-multimers}, \texttt{CASP14}, and \texttt{Benchmark2}. We perform 3 downsamplings using different seeds and report the mean and standard deviation.}
\vspace{-5pt}
\label{tab:sp_app}
\end{table}

\section{More Details for Protein Function Prediction and PPI Link Prediction} \label{app:graphdownstream}
We evaluate \themodelg's performance on two graph-related downstream tasks, protein structure prediction and PPI link prediction on OGB \citep{hu2020open}, i.e. \texttt{ogbn-proteins} and \texttt{ogbl-ppa}. Both datasets provide a PPI graph but with different scales (i.e. the numbers of nodes and edges) and predict different tasks (i.e. node property prediction and link prediction). Given the benefit of our graph-contextual training, we have a more principled way to obtain the protein representations from \themodelg or ESM-2-G by averaging the embeddings of \texttt{[BON]} and \texttt{[EON]} (Note that to impose the inductive bias of the learned distribution of positive random walk paths to the embedding, we concat an \texttt{[EDGE]} token right after \texttt{[EON]} such that the output embeddings will be pulled towards the positive, i.e. existing graph context). In contrast, we can only obtain the protein embeddings from ESM-2 or \themodel by averaging the AA token embeddings, which are not very informative. 
We show the evaluation results on \texttt{ogbn-proteins} in \Cref{tab:ogbn-proteins} against a set of popular baselines.
Our model achieves the best demonstration of the benefits of having graph context-aware protein embeddings learned in \themodelg. We also provide ablation studies in \Cref{tab:abl_ogbn,tab:abl_ogbl} where we choose two GNN backbones GCN and GraphSAGE to evaluate if the learned embeddings from \themodel and \themodelg are beneficial to this task. The evidence shows that the learned embeddings consistently improve the performance and the graph context provides another boost.

As discussed in \Cref{sec:con}, we can also perform more graph-specific self-supervised learning \citep{wang2021multi,wang2021molecular,wang2021molcloze,zhao2022graph} within the given graph context before supervised fine-tuning. By using this, we may obtain better initialization for node embeddings which would potentially encode the graph context better and improve the final prediction performance.


\begin{table}[ht]
    \centering
    \begin{minipage}{\linewidth}
        \centering
        \begin{tabular}{l c}
            \toprule
            \textbf{Model} & \textbf{Accuracy} \\
            \midrule
            Node2vec \citep{grover2016node2vec}        & 0.6881 $\pm$ 0.0065 \\
            GCN \citep{kipf2016semi}             & 0.7251 $\pm$ 0.0035 \\
            GraphSAGE \citep{hamilton2017inductive}       & 0.7443 $\pm$ 0.0064 \\
            DeepGCN \citep{li2019deepgcns}         & 0.8496 $\pm$ 0.0028 \\
            GAT \citep{velivckovic2017graph}            & 0.8501 $\pm$ 0.0046 \\
            DeeperGCN \citep{li2020deepergcn}     & 0.8580 $\pm$ 0.0017 \\
            UniMP \citep{shi2020masked}           & 0.8642 $\pm$ 0.0008 \\
            GIPA \citep{li2023gipa}            & 0.8700 $\pm$ 0.0010 \\
            \midrule
            ESM-2-G & {0.8920 $\pm$ 0.0008} \\
            \themodelg & \textbf{0.8925 $\pm$ 0.0010} \\
            \bottomrule
        \end{tabular}
        \caption{Performance on \texttt{ogbn-proteins}.}
        \label{tab:ogbn-proteins}
    \end{minipage}
\end{table}

\begin{table}[ht]
    \centering
    \begin{minipage}{0.45\linewidth}
        \centering
        \begin{tabular}{l c}
            \toprule
            \textbf{Model} & \textbf{Accuracy} \\
            \midrule
            GCN             & 0.7251 $\pm$ 0.0035 \\
            GCN+\themodel & {0.7643 $\pm$ 0.0042} \\
            GCN+\themodelg & \textbf{0.7668 $\pm$ 0.0056} \\
            \midrule
            GraphSAGE       & 0.7443 $\pm$ 0.0064 \\
            GraphSAGE + \themodel & {0.7662 $\pm$ 0.0021} \\
            GraphSAGE + \themodelg & \textbf{0.7679 $\pm$ 0.0029} \\
            \bottomrule
        \end{tabular}
        \caption{Ablations on \texttt{ogbn-proteins}.}
        \label{tab:abl_ogbn}
    \end{minipage}
    \hfill    
    \begin{minipage}{0.45\linewidth}
        \centering
        \begin{tabular}{l c}
            \toprule
            \textbf{Model} & \textbf{Hits@100} \\
            \midrule
            GCN             & 0.1867 $\pm$ 0.0132 \\
            GCN+\themodel & {0.1946 $\pm$ 0.0142} \\
            GCN+\themodelg & \textbf{0.1988 $\pm$ 0.0156} \\
            \midrule
            GraphSAGE       & 0.1655 $\pm$ 0.0240 \\
            GraphSAGE + \themodel & {0.1847 $\pm$ 0.0193} \\
            GraphSAGE + \themodelg & \textbf{0.1876 $\pm$ 0.0164} \\
            \bottomrule
        \end{tabular}
        \caption{Ablations on \texttt{ogbl-ppa}.}
        \label{tab:abl_ogbl}
    \end{minipage}%
\end{table}

\section{On the Absence of Massive Activations}
Recent studies \citep{sun2024massive} have highlighted the phenomenon of massive activations in Transformer-based models, particularly large language models (LLMs) such as LLaMA2 and GPT-2. These massive activations refer to a small number of hidden units exhibiting values that are several orders of magnitude larger than the median activation—often input-agnostic and acting as implicit bias terms throughout the model. This behavior, while seemingly innocuous, introduces challenges in quantization and numerical stability, and can negatively affect interpretability and robustness.

\begin{figure}
    \centering
    \includegraphics[width=\linewidth]{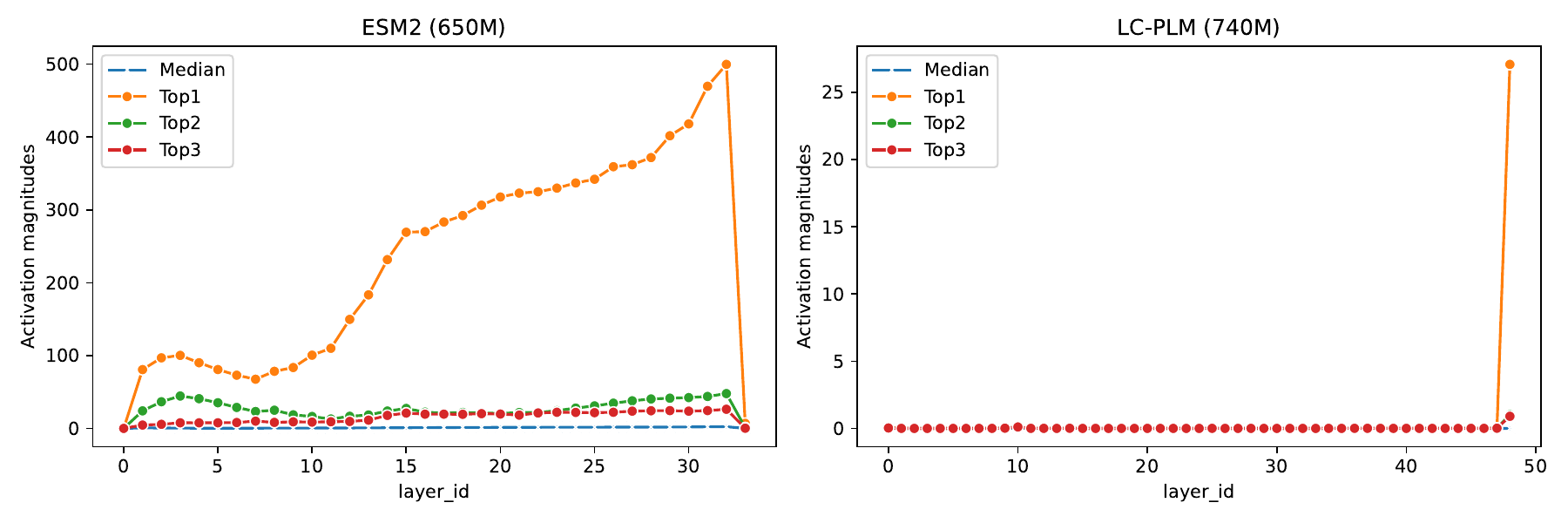}
    \caption{Activation magnitude of intermediate layers in ESM2 and \themodel.}
    \label{fig:massive_activation}
\end{figure}

To examine whether this issue extends to protein language models, we analyzed internal activations in ESM2 (650M) and \themodel (740M). Unlike ESM2, which clearly exhibits massive activations in intermediate layers (with magnitudes exceeding 400 in some layers), \themodel do not exhibit any such spikes in activation magnitude across layers (\Cref{fig:massive_activation}). Across all layers, the largest BiMamba activations remain within a modest range, closely tracking the median.

This absence of massive activations in \themodel has some implications: (1) Improved numerical stability: Lower variance in activation magnitudes reduces the risk of gradient explosion or instability during fine-tuning or continued pretraining; (2) Quantization-friendly: Models without large outlier activations are significantly easier to quantize using standard techniques, potentially allowing for more aggressive compression without accuracy loss. While the exact architectural mechanisms underlying the absence of massive activations in \themodel remain to be fully understood, we hypothesize that the recurrent state-space structure of Mamba layers inherently mitigates the formation of such activations.

\section{Pseudocode} \label{app:pseudo}
We provide a detailed breakdown of our algorithm in this section and then we summarize this computation procedure into a pseudocode algorithmic block as shown in \Cref{alg:pseudo}. The algorithm procedure can be stated as follows:
\begin{description}
\item[Input] $\mathbf{T}_{l-1}: (B, S, D)$ tensor, where $B$ is batch size, $S$ is sequence length, and $D$ is hidden state dimension.
\item[Output] $\mathbf{T}_{l}: (B, S, D)$ tensor.

\item[Process]
\begin{description}
    \item[1. Normalization] 
    $\mathbf{T}'_{l-1} : (B, S, D) \leftarrow \text{Norm}(\mathbf{T}_{l-1})$
    
    \item[2. Reversal] 
    $\hat{\mathbf{T}}'_{l-1} : (B, S, D) \leftarrow \text{Reverse}\left(\mathbf{T}'_{l-1}\right)$
    
    \item[3. Parallel Processing] For both $\mathbf{T}'_{l-1}$ and $\hat{\mathbf{T}}'_{l-1}$, denoted as $\mathbf{T}'_{*,l-1}$:
    \begin{description}
        \item[a. Linear Transformations] 
        \begin{align*}
        \mathbf{X}_{*,l-1}: (B, S, E) &\leftarrow \text{Linear}^{\mathbf{X}_{*,l-1}}(\mathbf{T}'_{*,l-1}) \\
        \mathbf{Z}_{*,l-1}: (B, S, E) &\leftarrow \text{Linear}^{\mathbf{Z}_{*,l-1}}(\mathbf{T}'_{*,l-1})
        \end{align*}
        
        \item[b. Convolution and Activation] 
        $\mathbf{X}'_{*,l-1} : (B, S, E) \leftarrow \text{SiLU}(\text{Conv1d}(\mathbf{X}_{*,l-1}))$
        
        \item[c. Additional Linear Transformations] 
        \begin{align*}
        \mathbf{B}_{*,l-1}: (B, S, N) &\leftarrow \text{Linear}^{B}_{*,l-1}(\mathbf{X}'_{*,l-1}) \\
        \mathbf{C}_{*,l-1}: (B, S, N) &\leftarrow \text{Linear}^{C}_{*,l-1}(\mathbf{X}'_{*,l-1})
        \end{align*}
        
        \item[d. Delta Computation] 
       \begin{align*}\Delta_{*,l-1} : (B, S, E) \leftarrow \log(1 + \exp(\text{Linear}_{*,l-1}^{\Delta}(\mathbf{X}'_{*,l-1}) + \text{Parameter}^{\Delta}_{*,l-1}))\end{align*}
        
        \item[e. Parameter Scaling] 
        $\bar{\mathbf{A}}_{*,l-1} : (B, S, E, N) \leftarrow \Delta_{*,l-1} \otimes \text{Parameter}^{\bar{A}}_{*,l-1}$
        
        \item[f. B Update] 
        $\mathbf{B}_{*,l-1} : (B, S, E, N) \leftarrow \Delta_{*,l-1} \otimes \mathbf{B}_{*,l-1}$
        
        \item[g. State Space Model Application] 
        \begin{align*}\mathbf{Y}_{*,l-1} : (B, S, E) \leftarrow \text{SSM}(\bar{\mathbf{A}}_{*,l-1}, \mathbf{B}_{*,l-1}, \mathbf{C}_{*,l-1})(\mathbf{X}'_{*,l-1})\end{align*}
        
        \item[h. Final Computation] 
        $\mathbf{Y}'_{*,l-1} : (B, S, E) \leftarrow \mathbf{Y}_{*,l-1} \odot \text{SiLU}(\mathbf{Z}_{*,l-1})$
    \end{description}
    
    \item[4. Combination and Output] 
    $\mathbf{T}_{l} : (B, S, D) \leftarrow \text{Linear}^{T}(\mathbf{Y}'_{l-1} + \hat{\mathbf{Y}}'_{l-1}) + \mathbf{T}_{l-1}$
\end{description}

\item[Key Operations]
\begin{description}
    \item[Norm] Normalization operation
    \item[Linear] Linear transformation
    \item[Conv1d] 1D Convolution
    \item[SiLU] Sigmoid Linear Unit activation function
    \item[SSM] State Space Model
    \item[$\otimes$] Element-wise multiplication
    \item[$\odot$] Element-wise multiplication
\end{description}
\end{description}

\begin{algorithm}
\caption{\themodels Block}\label{alg:pseudo}
\textbf{Input:} $\mathbf{T}_{l-1}: (B, S, D)$ \\
\textbf{Output:} $\mathbf{T}_{l}: (B, S, D)$
\begin{algorithmic}[1]
\State $\mathbf{T}'_{l-1} : (B, S, D) \leftarrow \text{Norm}(\mathbf{T}_{l-1})$
\State $\hat{\mathbf{T}}'_{l-1} : (B, S, D) \leftarrow \text{Reverse}\left(\mathbf{T}'_{l-1}\right)$
\For{$\mathbf{T}'_{*,l-1} \in \{\mathbf{T}'_{l-1}, \hat{\mathbf{T}}'_{l-1}\}$}
    \State $\mathbf{X}_{*,l-1}: (B, S, E) \leftarrow \text{Linear}^{\mathbf{X}_{*,l-1}}(\mathbf{T}'_{*,l-1})$
    \State $\mathbf{Z}_{*,l-1}: (B, S, E) \leftarrow \text{Linear}^{\mathbf{Z}_{*,l-1}}(\mathbf{T}'_{*,l-1})$
    \State $\mathbf{X}'_{*,l-1} : (B, S, E) \leftarrow \text{SiLU}(\text{Conv1d}(\mathbf{X}_{*,l-1}))$
    \State $\mathbf{B}_{*,l-1}: (B, S, N) \leftarrow \text{Linear}^{B}_{*,l-1}(\mathbf{X}'_{*,l-1})$
    \State $\mathbf{C}_{*,l-1}: (B, S, N) \leftarrow \text{Linear}^{C}_{*,l-1}(\mathbf{X}'_{*,l-1})$
    \State $\Delta_{*,l-1} : (B, S, E) \leftarrow \log(1 + \exp(\text{Linear}_{*,l-1}^{\Delta}(\mathbf{X}'_{*,l-1}) + \text{Parameter}^{\Delta}_{*,l-1}))$
    \State $\bar{\mathbf{A}}_{*,l-1} : (B, S, E, N) \leftarrow \Delta_{*,l-1} \otimes \text{Parameter}^{\bar{A}}_{*,l-1}$
    \State $\mathbf{B}_{*,l-1} : (B, S, E, N) \leftarrow \Delta_{*,l-1} \otimes \mathbf{B}_{*,l-1}$
    \State $\mathbf{Y}_{*,l-1} : (B, S, E) \leftarrow \text{SSM}(\bar{\mathbf{A}}_{*,l-1}, \mathbf{B}_{*,l-1}, \mathbf{C}_{*,l-1})(\mathbf{X}'_{*,l-1})$
    \State $\mathbf{Y}'_{*,l-1} : (B, S, E) \leftarrow \mathbf{Y}_{*,l-1} \odot \text{SiLU}(\mathbf{Z}_{*,l-1})$
\EndFor
\State $\mathbf{T}_{l} : (B, S, D) \leftarrow \text{Linear}^{T}(\mathbf{Y}'_{l-1} + \hat{\mathbf{Y}}'_{l-1}) + \mathbf{T}_{l-1}$
\State \textbf{return} $\mathbf{T}_{l}$
\end{algorithmic}
\end{algorithm}

\section{Sampling Bias in Random Walk}\label{app:sampling}
For random walk sampling, there are two parameters $p$ and $q$ we can use to control the direction of exploration (a balance between the depth-first search (DFS) and breath-first search (BFS)).

\paragraph{Return Parameter $p$}

Parameter $p$ controls the likelihood of immediately revisiting a node. A high value ($p > \max(q, 1)$) reduces the chance of sampling an already-visited node, encouraging moderate exploration and avoiding 2-hop redundancy.

\paragraph{In-out Parameter $q$}

Parameter $q$ allows the search to differentiate between "inward" and "outward" nodes:
\begin{itemize}
    \item If $q > 1$, the walk is biased towards nodes close to node $t$, approximating BFS behavior.
    \item If $q < 1$, the walk tends to visit nodes further from node $t$, encouraging outward exploration similar to DFS.
\end{itemize}

\paragraph{Benefits over Pure BFS/DFS}

Random walks offer several advantages over traditional BFS/DFS approaches:

\begin{enumerate}
    \item \textbf{Computational Efficiency:} They are efficient in terms of both space and time requirements.
    \item \textbf{Scalability:} The space complexity to explore the immediate neighbors of every node is $O(|E|)$ for a graph with edge set $E$.
    \item \textbf{Flexibility:} For 2nd order random walks, the space complexity becomes $O(a^2|V|)$, where $a$ is the average degree of the graph and $V$ is the vertex set.
    \item \textbf{Effective Sampling:} Random walks provide a convenient mechanism to increase the effective sampling rate by reusing samples across different source nodes.
    \item \textbf{Parallel Sampling:} Due to the Markovian nature of the random walk, $k$ samples for $l - k$ nodes can be generated at once, resulting in an effective complexity of $O(\frac{l}{k(l-k)})$ per sample.
\end{enumerate}

This approach combines the benefits of BFS and DFS, allowing for a more nuanced exploration of network structures that exhibit both structural equivalence and homophily.

\section{Training and Evaluation Curves}

\subsection{The First-stage Pretraining}
\begin{figure}[h!]
    \centering
    \begin{subfigure}{0.45\linewidth}
        \centering
        \includegraphics[width=\linewidth]{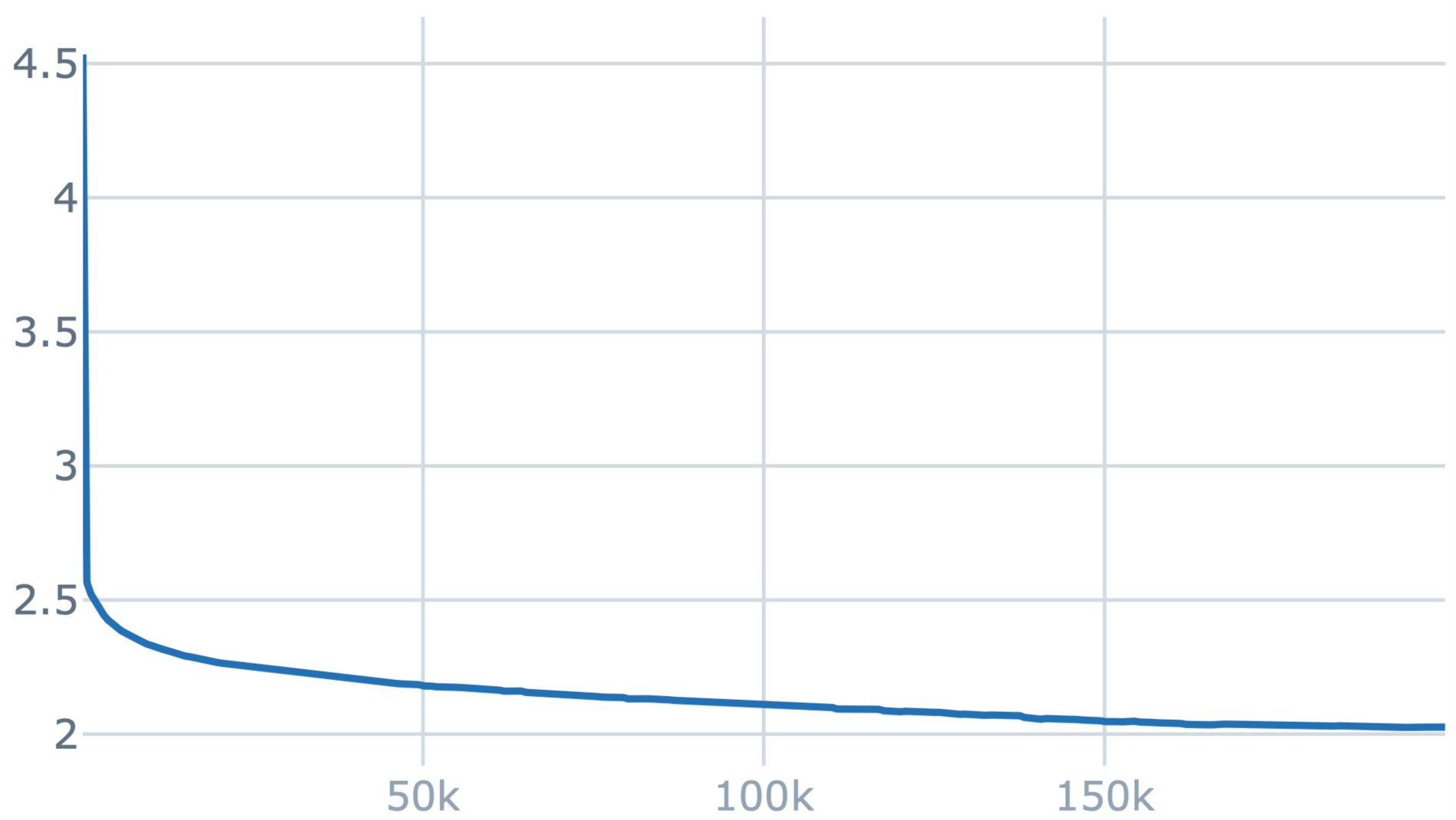}
        \caption{Training curve.}
        \label{fig:log_100b_train}
    \end{subfigure}
    \hfill
    \begin{subfigure}{0.45\linewidth}
        \centering
        \includegraphics[width=\linewidth]{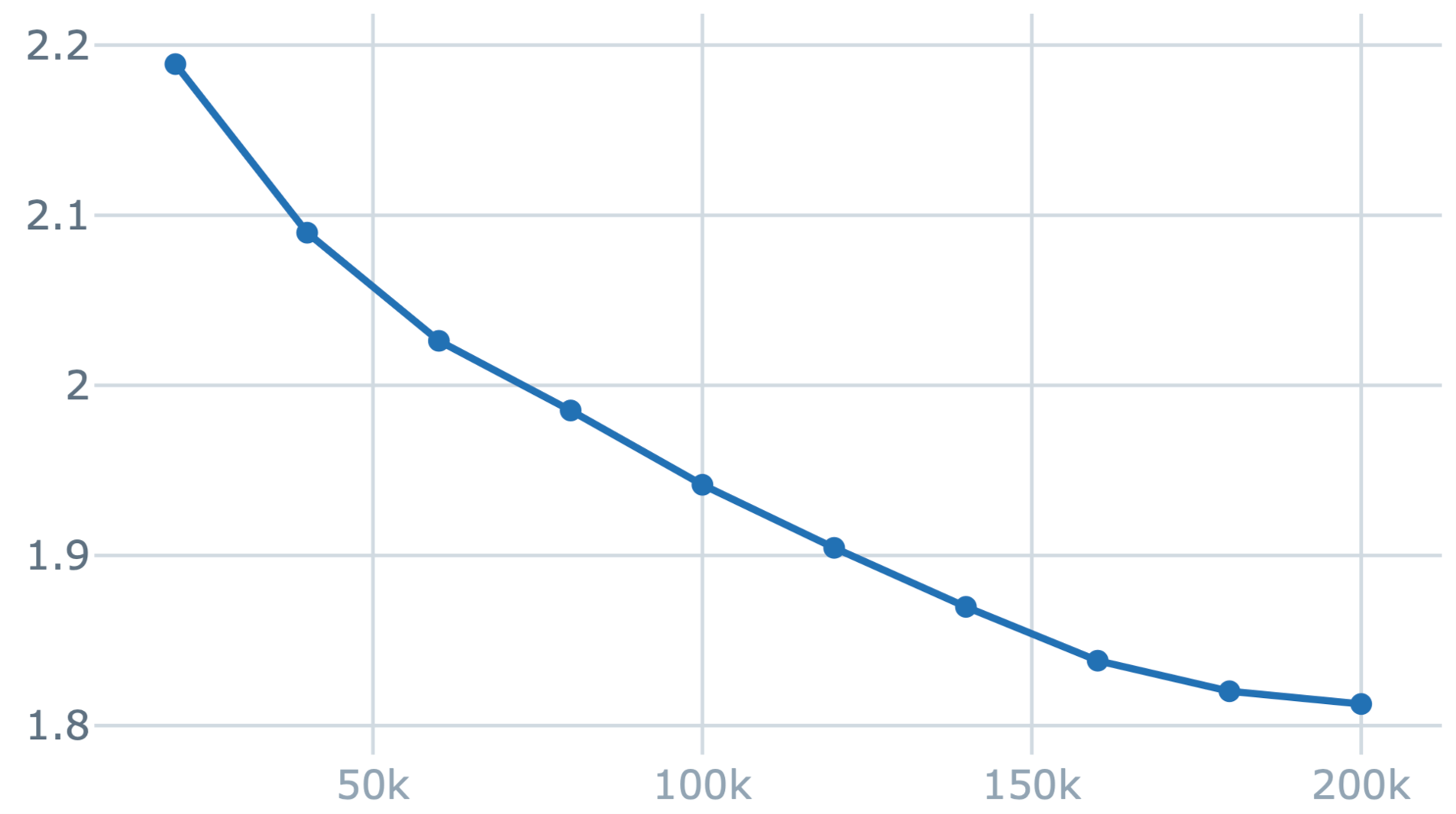}
        \caption{Evaluation curve.}
        \label{fig:log_100b_eval}
    \end{subfigure}
    \caption{The first-stage pretraining of 740M \themodel on 100B \texttt{UniRef50}. The evaluation set is 250K \texttt{UniRef90}.}
    \label{fig:log_100b}
\end{figure}

We show the training and evaluation loss curves in \Cref{fig:log_100b} for our first-stage pretraining of \themodel on 100B \texttt{UniRef50}. This pretrained model is used in all downstream task evaluations reported in the paper.

\subsection{The Scaling Law Experiments}
\begin{figure}[h!]
    \centering
    \begin{subfigure}{0.49\linewidth}
        \centering
        \includegraphics[width=\linewidth]{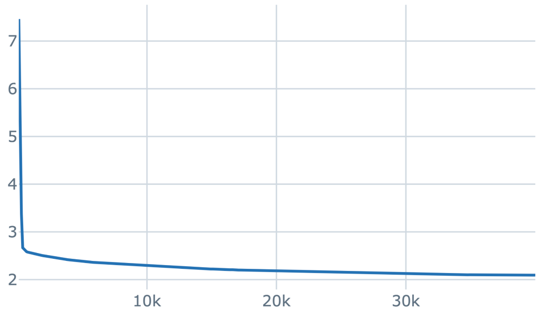}
        \caption{Training curve.}
        \label{fig:log_scaling_130_train}
    \end{subfigure}
    \hfill
    \begin{subfigure}{0.49\linewidth}
        \centering
        \includegraphics[width=\linewidth]{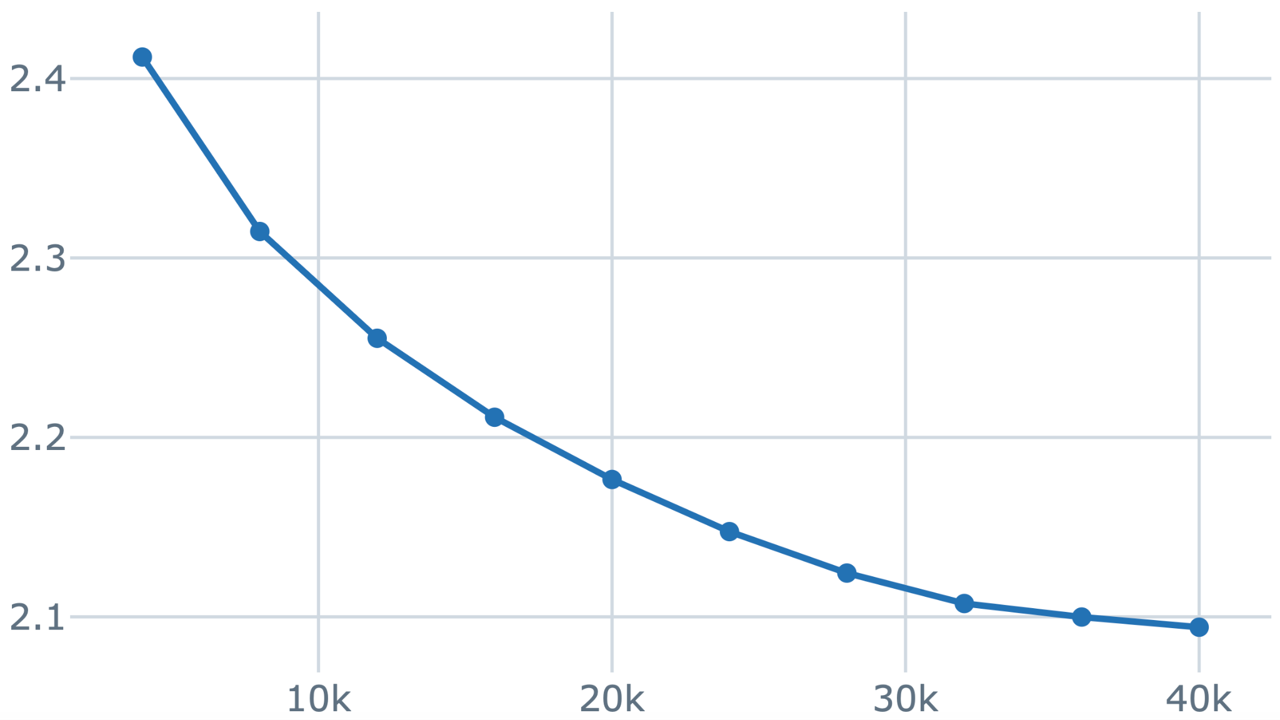}
        \caption{Evaluation curve.}
        \label{fig:log_scaling_130_eval}
    \end{subfigure}
    \caption{The scaling law training of 100M \themodel on 20B \texttt{UniRef90}. The evaluation set is 250K \texttt{UniRef90}.}
    \label{fig:log_scaling_130}
\end{figure}

\begin{figure}[h!]
    \centering
    \begin{subfigure}{0.49\linewidth}
        \centering
        \includegraphics[width=\linewidth]{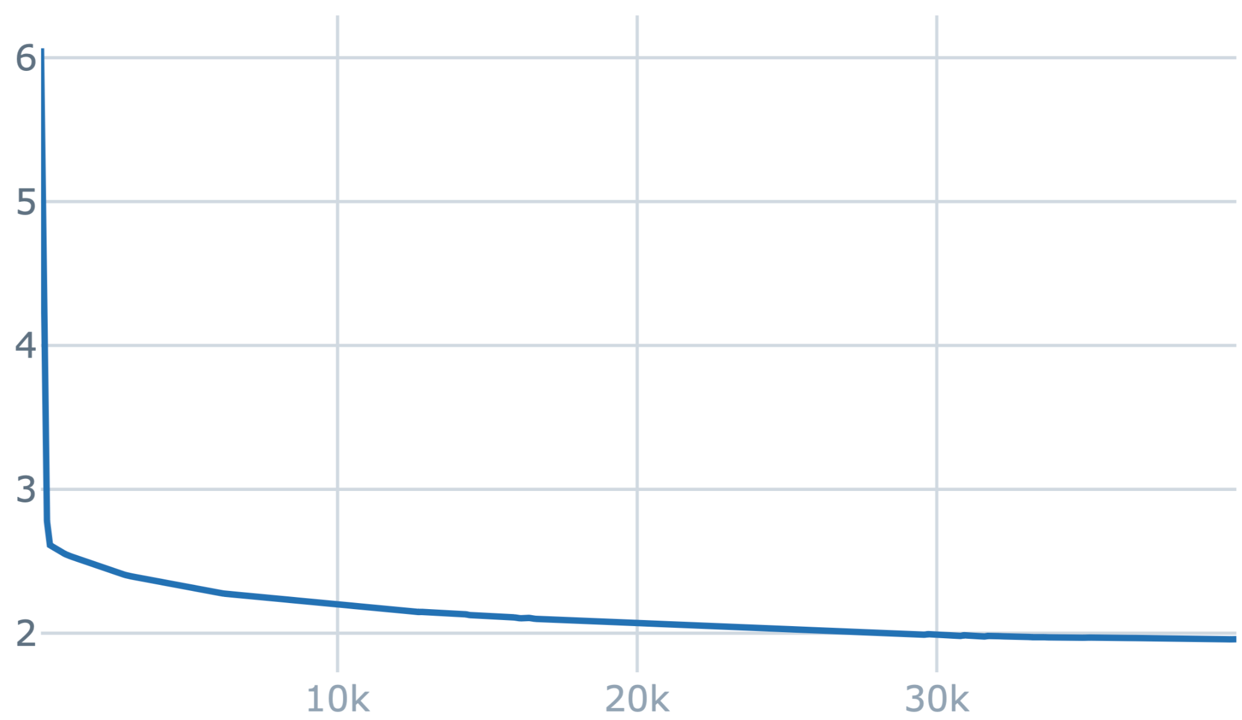}
        \caption{Training curve.}
        \label{fig:log_scaling_370_train}
    \end{subfigure}
    \hfill
    \begin{subfigure}{0.49\linewidth}
        \centering
        \includegraphics[width=\linewidth]{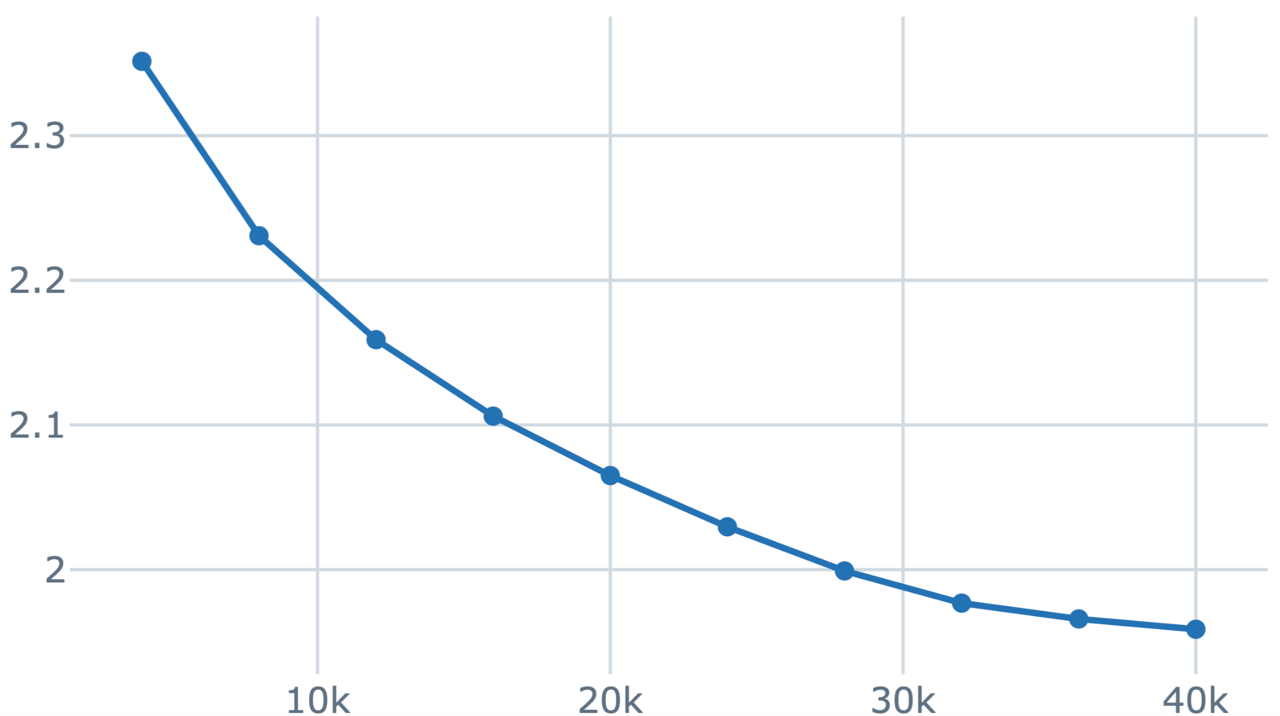}
        \caption{Evaluation curve.}
        \label{fig:log_scaling_370_eval}
    \end{subfigure}
    \caption{The scaling law training of 340M \themodel on 20B \texttt{UniRef90}. The evaluation set is 250K \texttt{UniRef90}.}
    \label{fig:log_scaling_370}
    \vspace{-10pt}
\end{figure}

\begin{figure}[h!]
    \centering
    \begin{subfigure}{0.49\linewidth}
        \centering
        \includegraphics[width=\linewidth]{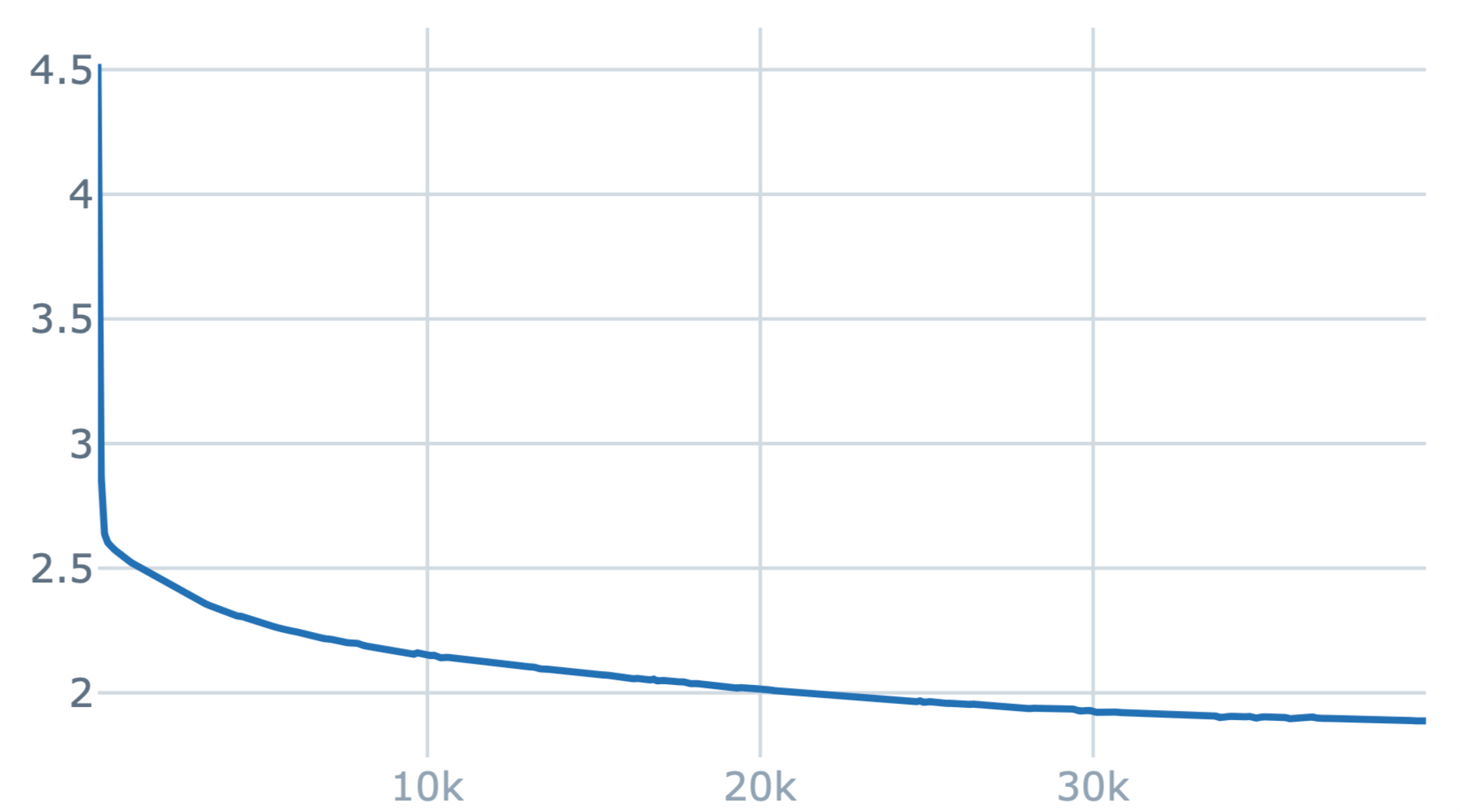}
        \caption{Training curve.}
        \label{fig:log_scaling_790_train}
    \end{subfigure}
    \hfill
    \begin{subfigure}{0.49\linewidth}
        \centering
        \includegraphics[width=\linewidth]{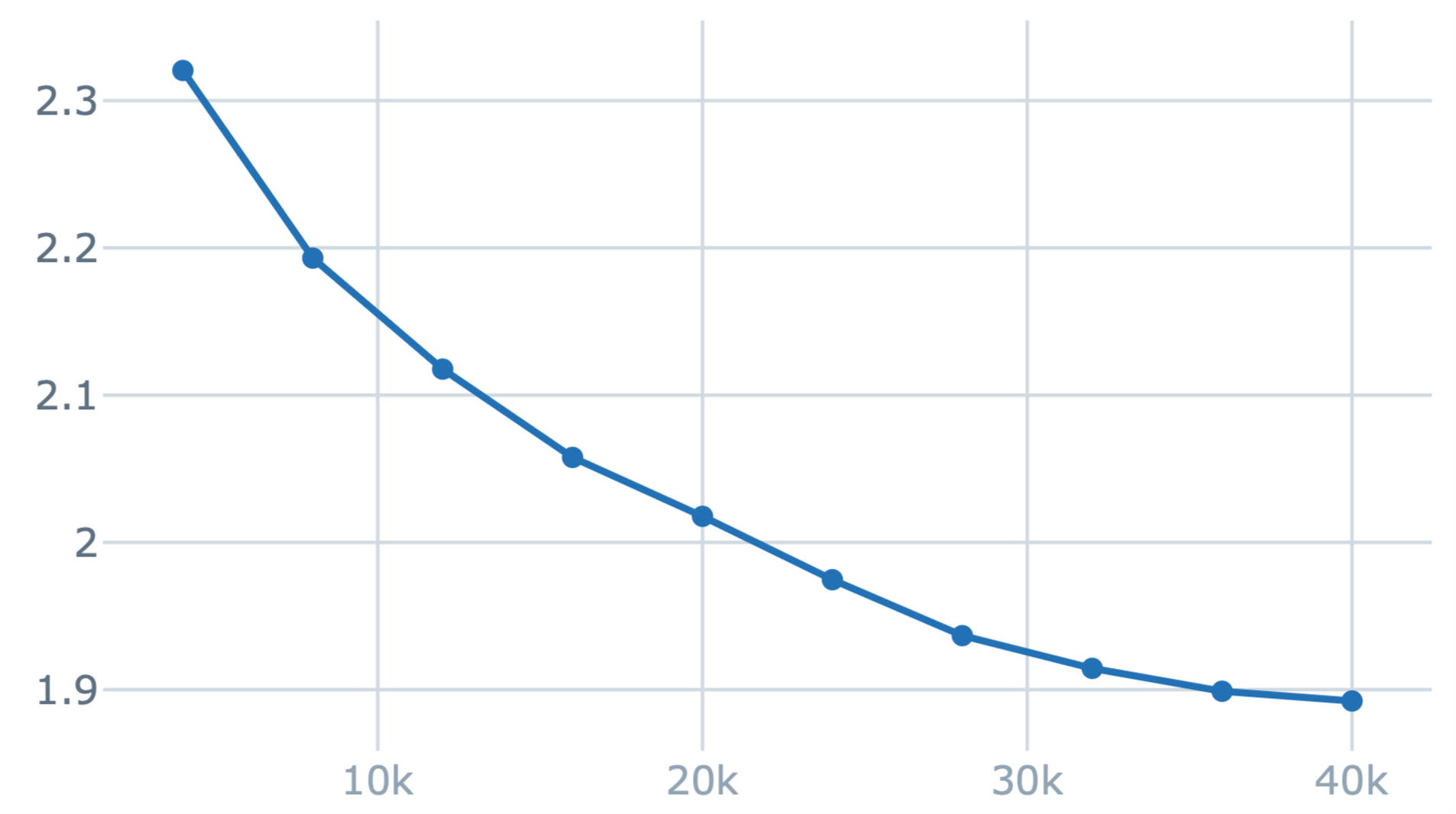}
        \caption{Evaluation curve.}
        \label{fig:log_scaling_790_eval}
    \end{subfigure}
    \caption{The scaling law training of 740M \themodel on 20B \texttt{UniRef90}. The evaluation set is 250K \texttt{UniRef90}.}
    \label{fig:log_scaling_790}
    \vspace{-10pt}
\end{figure}

\begin{figure}[h!]
    \centering
    \begin{subfigure}{0.49\linewidth}
        \centering
        \includegraphics[width=\linewidth]{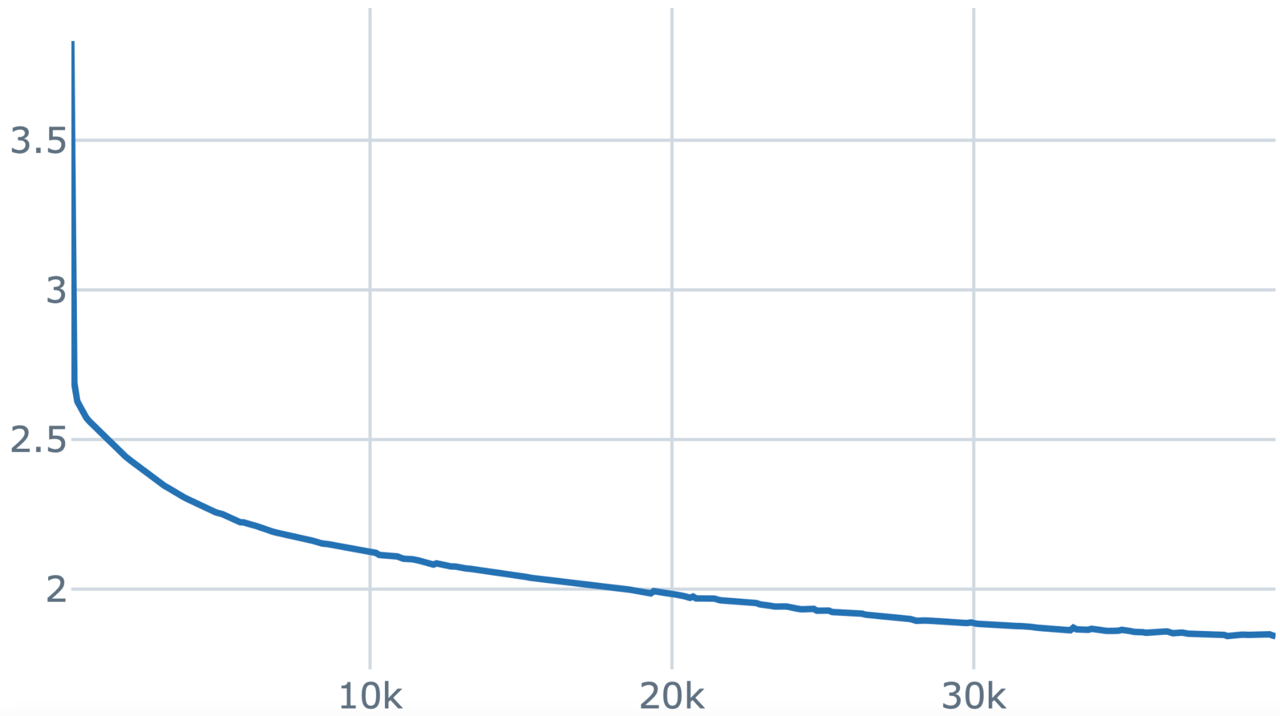}
        \caption{Training curve.}
        \label{fig:log_scaling_14b_train}
    \end{subfigure}
    \hfill
    \begin{subfigure}{0.49\linewidth}
        \centering
        \includegraphics[width=\linewidth]{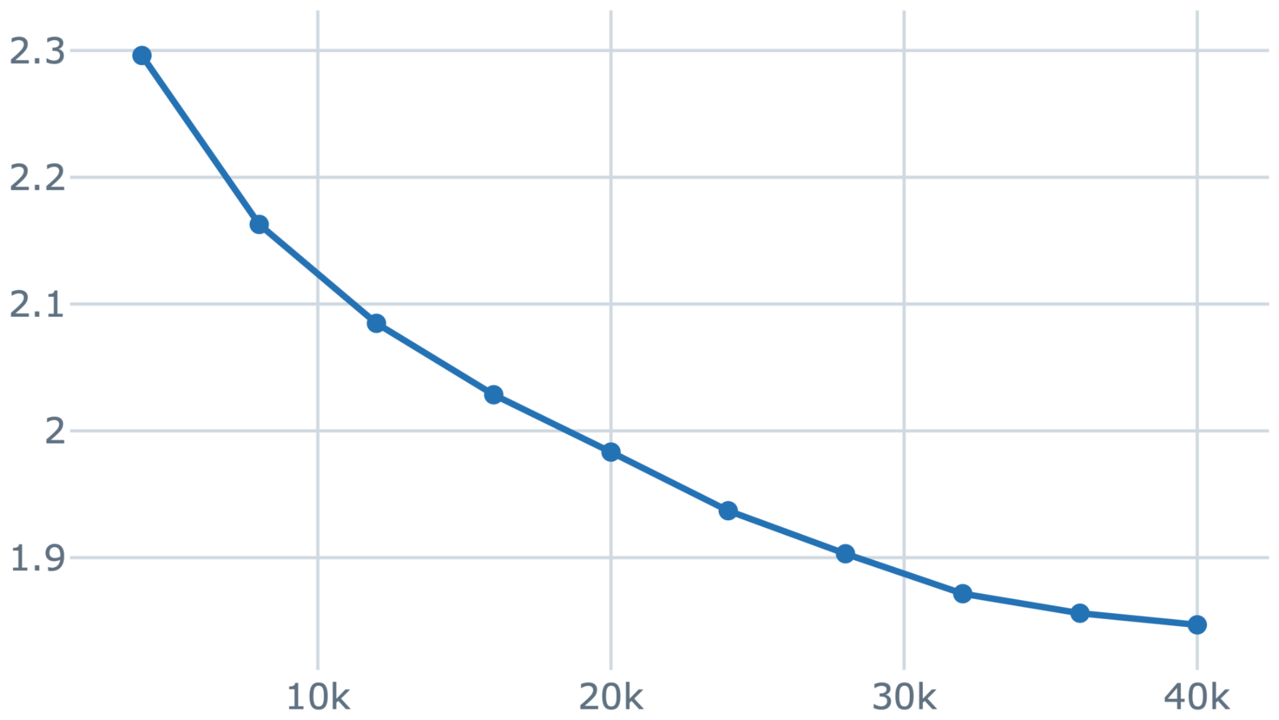}
        \caption{Evaluation curve.}
        \label{fig:log_scaling_14b_eval}
    \end{subfigure}
    \caption{The scaling law training of 1.3B \themodel on 20B \texttt{UniRef90}. The evaluation set is 250K \texttt{UniRef90}.}
    \label{fig:log_scaling_14b}
    \vspace{-10pt}
\end{figure}

We show the training and evaluation loss curves in \Cref{fig:log_scaling_130}, \Cref{fig:log_scaling_370}, \Cref{fig:log_scaling_790}, and \Cref{fig:log_scaling_14b} for our scaling law training of 100M, 340M, 740M, and 1.3B \themodel on 20B \texttt{UniRef90}. The evaluation set is the held-out 250K \texttt{UniRef90}.

\subsection{The Length Extrapolation Experiments}
\begin{figure}[h!]
    \centering
    \begin{subfigure}{0.49\linewidth}
        \centering
        \includegraphics[width=\linewidth]{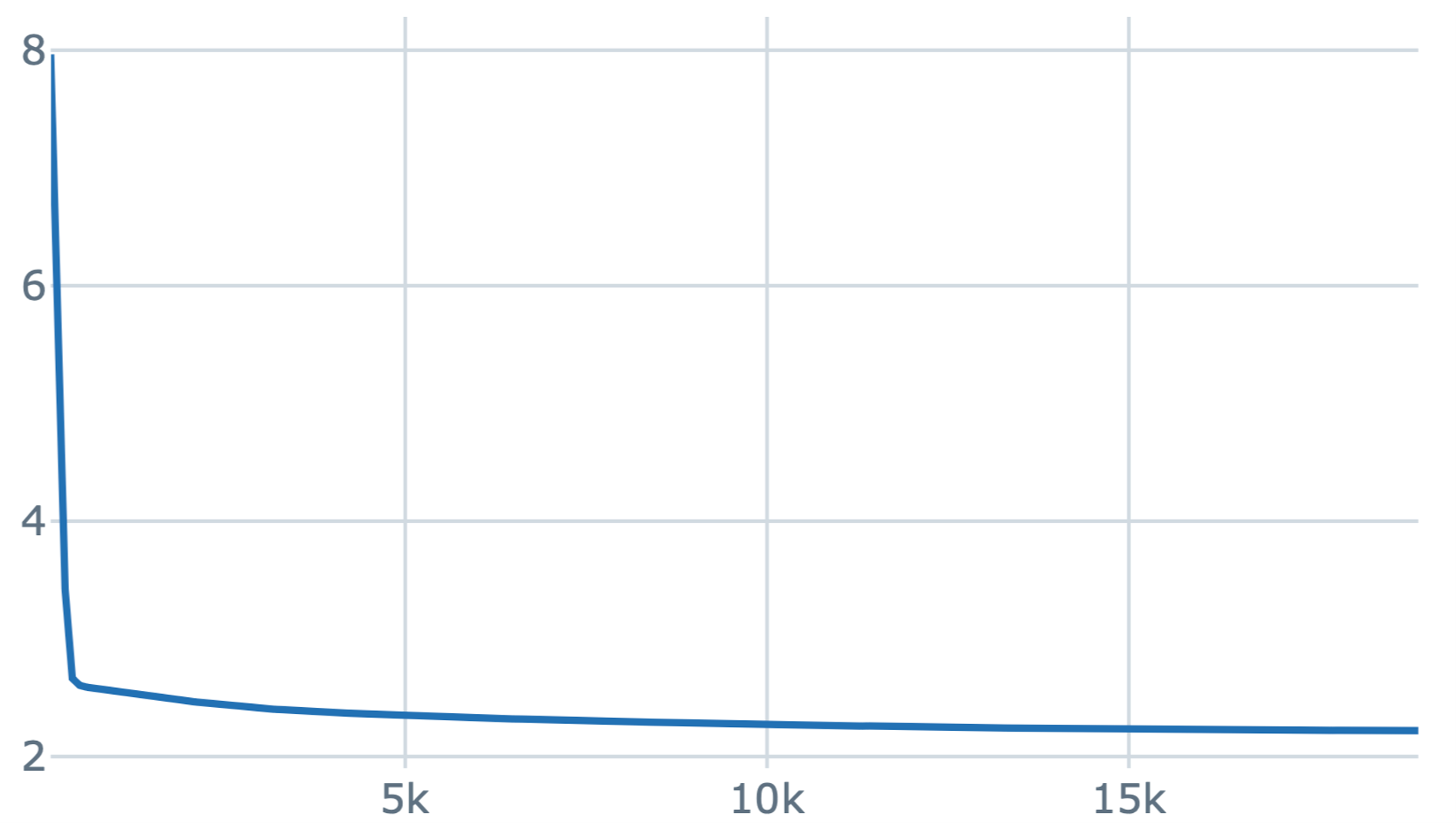}
        \caption{Training curve.}
        \label{fig:log_len_extra_130_train}
    \end{subfigure}
    \hfill
    \begin{subfigure}{0.49\linewidth}
        \centering
        \includegraphics[width=\linewidth]{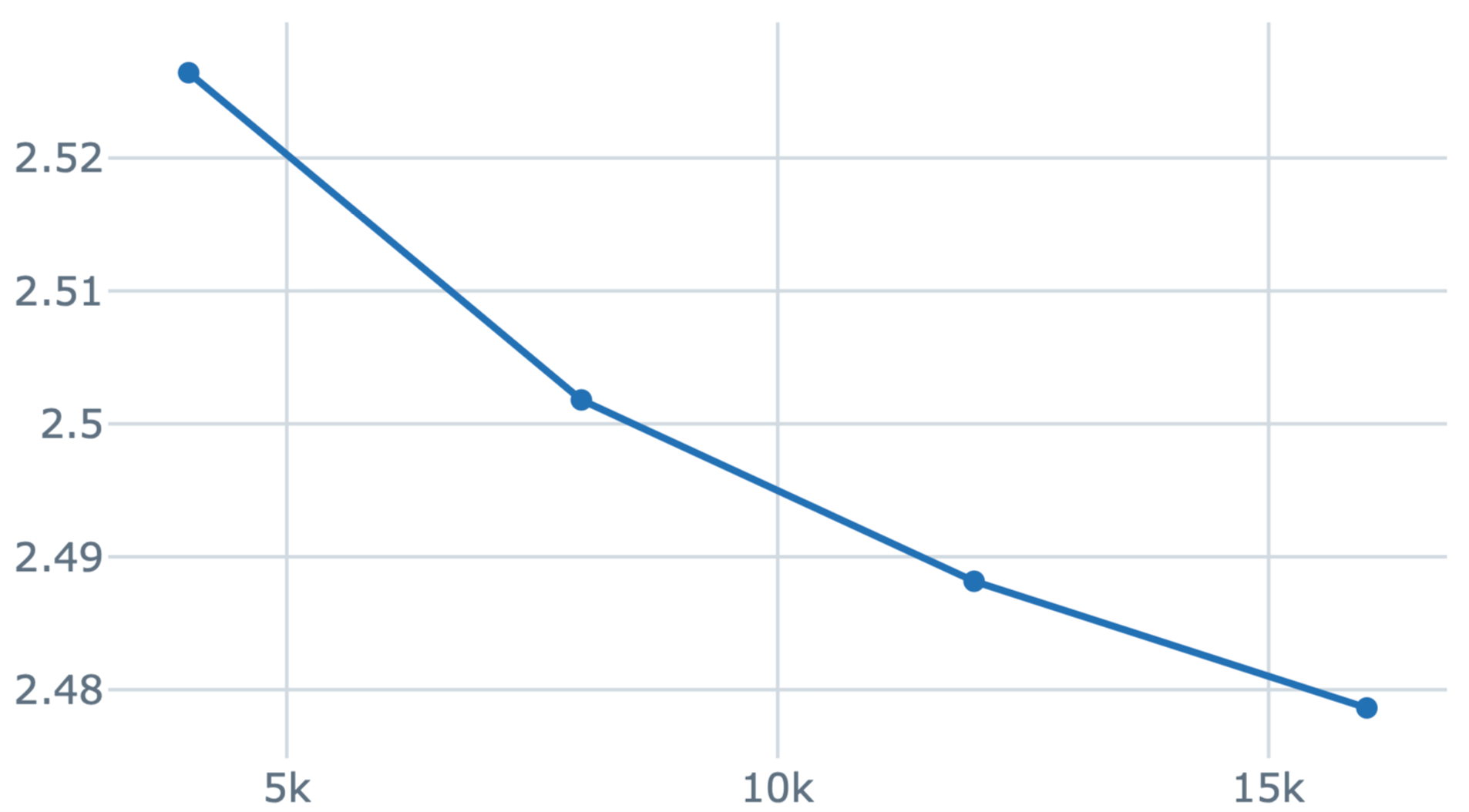}
        \caption{Evaluation curve.}
        \label{fig:log_len_extra_130_eval}
    \end{subfigure}
    \caption{The length extrapolation training of 100M \themodel on the 128-256 bin of \texttt{UniRef90}. The evaluation set is 250K \texttt{UniRef90}.}
    \label{fig:log_len_extra_130}
    \vspace{-10pt}
\end{figure}

\begin{figure}[h!]
    \centering
    \begin{subfigure}{0.49\linewidth}
        \centering
        \includegraphics[width=\linewidth]{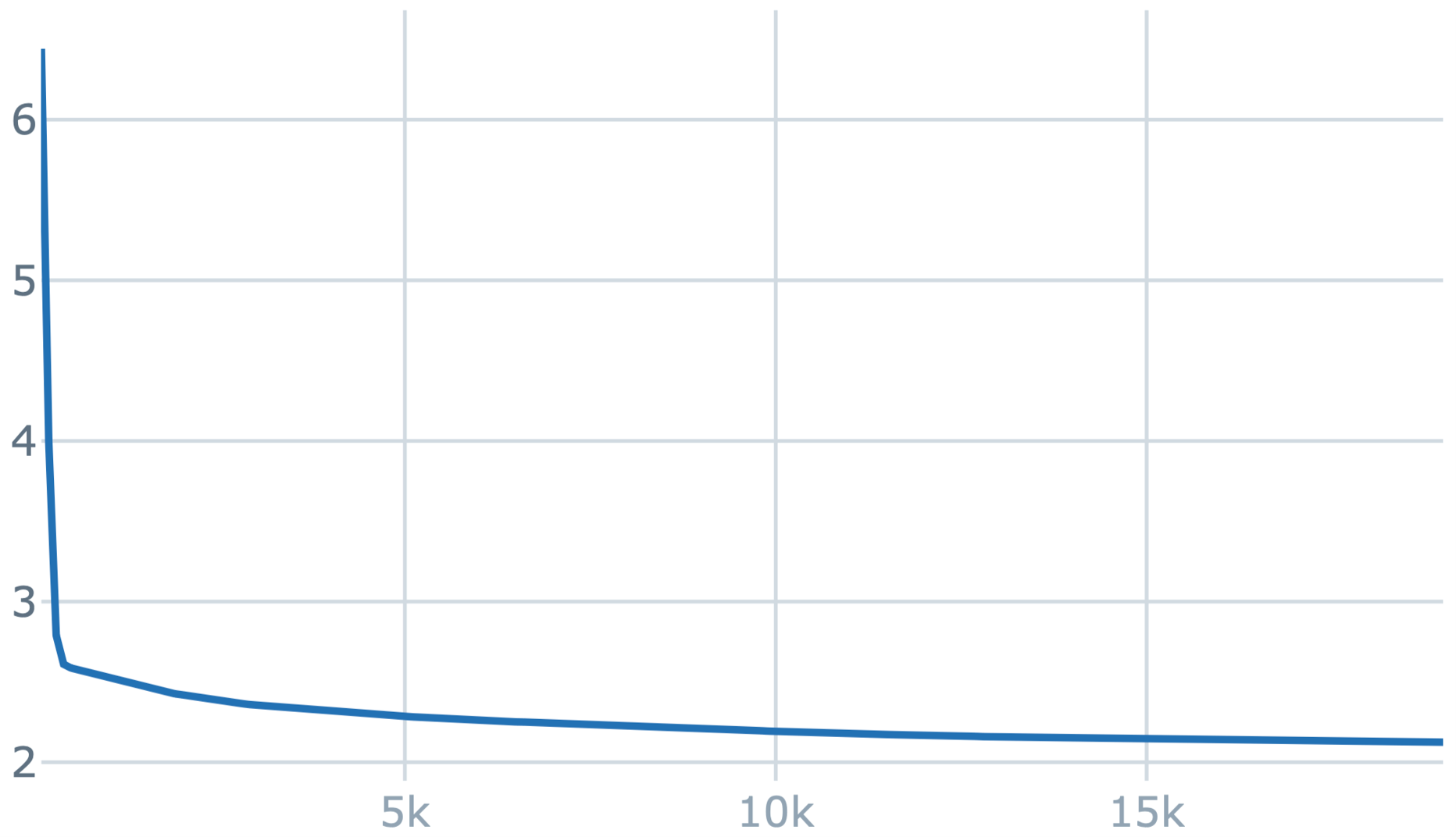}
        \caption{Training curve.}
        \label{fig:log_len_extra_370_train}
    \end{subfigure}
    \hfill
    \begin{subfigure}{0.49\linewidth}
        \centering
        \includegraphics[width=\linewidth]{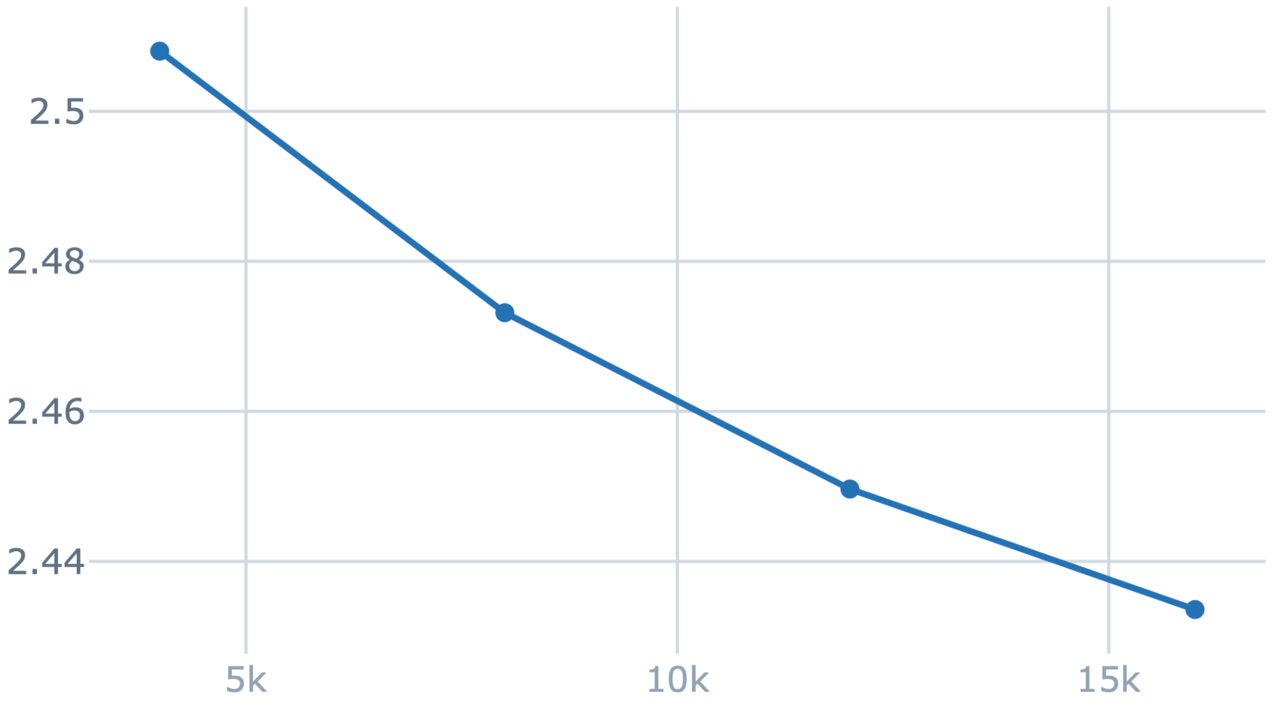}
        \caption{Evaluation curve.}
        \label{fig:log_len_extra_370_eval}
    \end{subfigure}
    \caption{The length extrapolation training of 340M \themodel on the 128-256 bin of \texttt{UniRef90}. The evaluation set is 250K \texttt{UniRef90}.}
    \label{fig:log_len_extra_370}
    \vspace{-10pt}
\end{figure}

\begin{figure}[h!]
    \centering
    \begin{subfigure}{0.49\linewidth}
        \centering
        \includegraphics[width=\linewidth]{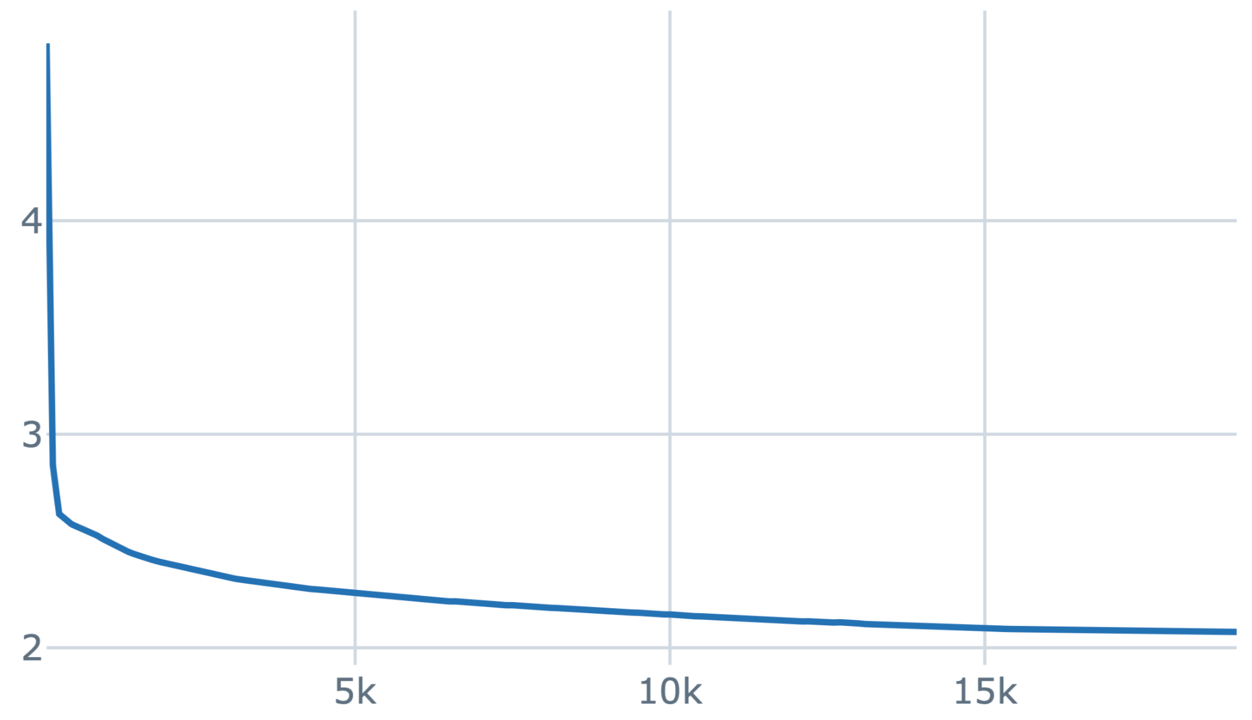}
        \caption{Training curve.}
        \label{fig:log_len_extra_790_train}
    \end{subfigure}
    \hfill
    \begin{subfigure}{0.49\linewidth}
        \centering
        \includegraphics[width=\linewidth]{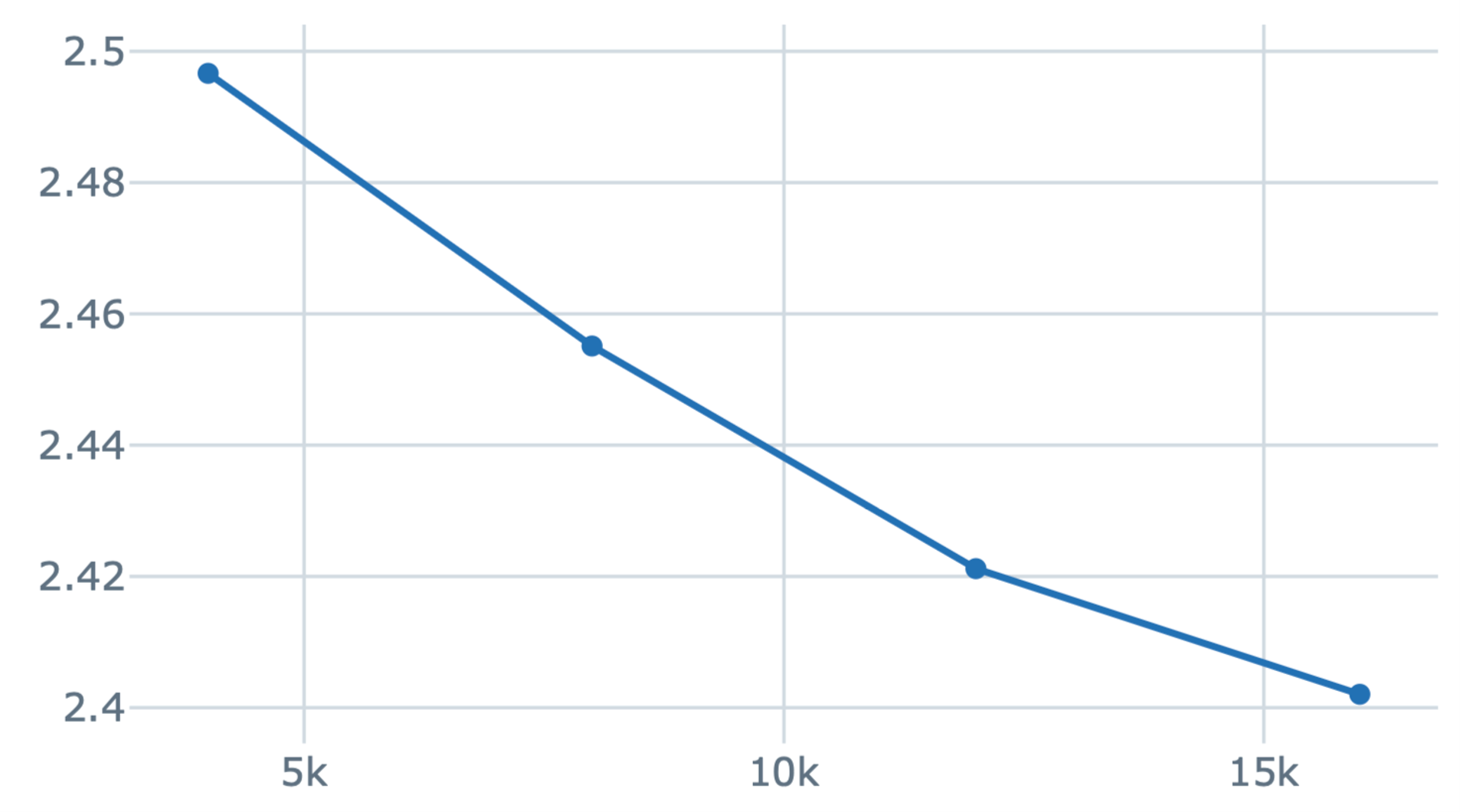}
        \caption{Evaluation curve.}
        \label{fig:log_len_extra_790_eval}
    \end{subfigure}
    \caption{The length extrapolation training of 740M \themodel on the 128-256 bin of \texttt{UniRef90}. The evaluation set is 250K \texttt{UniRef90}.}
    \label{fig:log_len_extra_790}
    \vspace{-10pt}
\end{figure}

We show the training and evaluation loss curves in \Cref{fig:log_len_extra_130}, \Cref{fig:log_len_extra_370}, and \Cref{fig:log_len_extra_790} for our scaling law training of 100M, 340M, and 740M \themodel on the 128-256 bin of \texttt{UniRef90}. The evaluation set is the held-out 250K \texttt{UniRef90}.

\subsection{The Second-stage Graph Contextual Training}
\begin{figure}[h!]
    \centering
    \begin{subfigure}{0.49\linewidth}
        \centering
        \includegraphics[width=\linewidth]{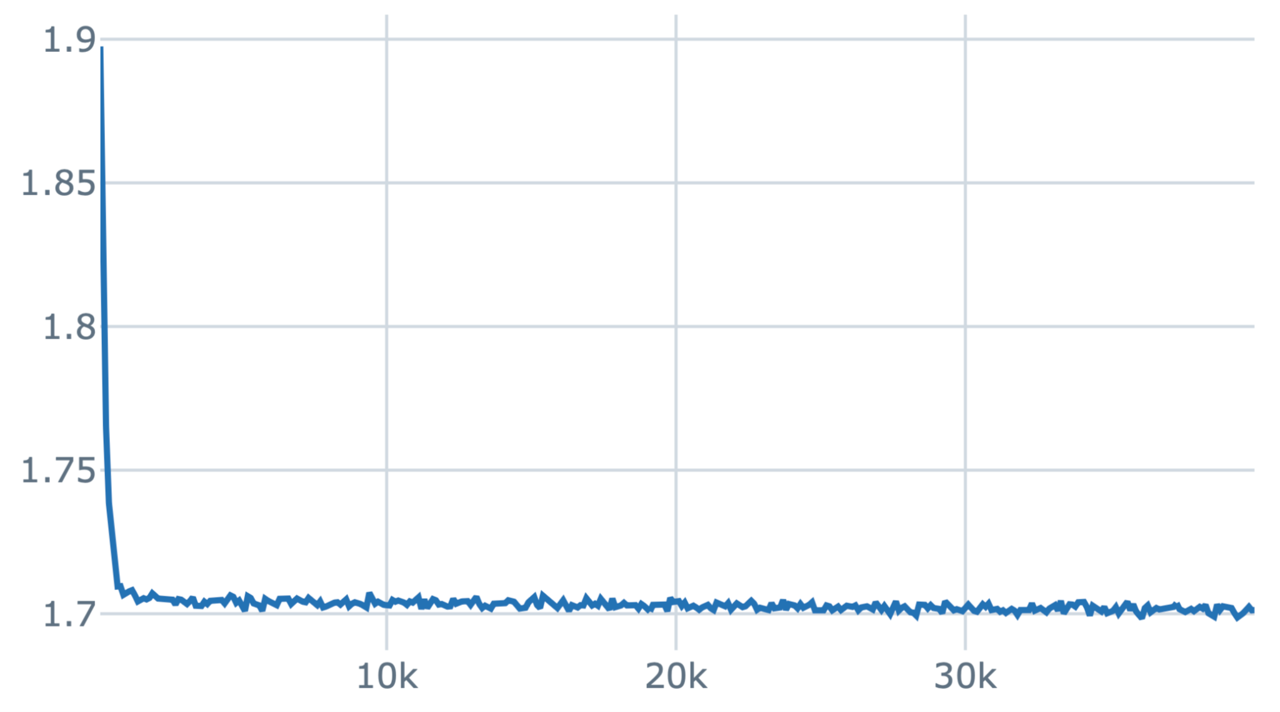}
        \caption{Training curve on \texttt{ogbn-proteins}.}
        \label{fig:log_proteins}
    \end{subfigure}
    \hfill
    \begin{subfigure}{0.49\linewidth}
        \centering
        \includegraphics[width=\linewidth]{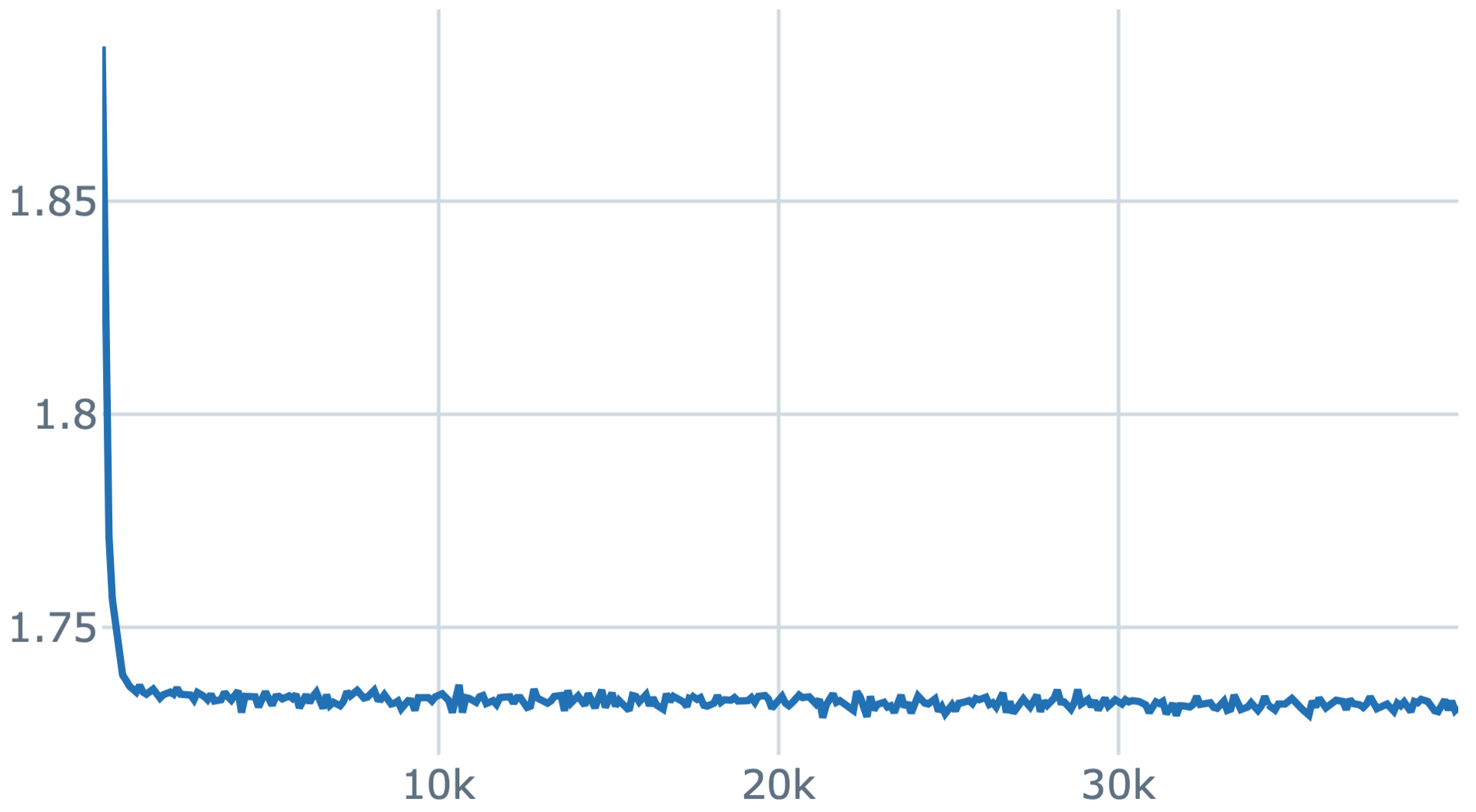}
        \caption{Training curve on \texttt{ogbl-ppa}.}
        \label{fig:log_ppa}
    \end{subfigure}
    \caption{The second-stage training of 740M \themodelg on \texttt{ogbn-proteins} and \texttt{ogbl-ppa}. The evaluation set is 250K \texttt{UniRef90}.}
    \label{fig:log_graph}
    \vspace{-10pt}
\end{figure}

We show the training loss curves in \Cref{fig:log_proteins}, and \Cref{fig:log_ppa} for our second-stage graph contextual training of 740M \themodelg on protein sequences included in \texttt{ogbn-proteins} and \texttt{ogbl-ppa}. The evaluation set is the held-out 250K \texttt{UniRef90}.

\end{appendices}

\end{document}